\documentclass[a4paper,11pt]{article}
\usepackage{jcappub} % for details on the use of the package, please
\usepackage[T1]{fontenc} % if needed
\usepackage{float} 
\usepackage{lmodern}
\usepackage{booktabs}
\usepackage{siunitx}
\usepackage[english]{babel}
\addto\captionsenglish{
  \renewcommand{\figurename}{Figure}
  \renewcommand{\tablename}{Table}
}
\usepackage[utf8]{inputenc}
\usepackage{natbib}
\usepackage[colorlinks=true, citecolor=blue, urlcolor=blue, linkcolor=blue]{hyperref} 
\usepackage{graphicx}
\usepackage{subfigure}
\usepackage[justification=raggedright]{caption}
\usepackage{tabularx}
\usepackage{dcolumn}
\usepackage{bm}
\usepackage{xcolor}

\definecolor{revcolor}{rgb}{0.0, 0.4, 0.7} 

\title{Inferring neutron-star Love-Q relations from gravitational waves in the
hierarchical Bayesian framework}

\author[a]{Zhihao Zheng,}
\author[b,c,1]{Ziming Wang\note{Corresponding authors.},}
\author[d]{Jinwen Deng,}
\author[b,c]{Yiming Dong,}
\author[c,e,1]{and Lijing Shao}

\affiliation[a]{School of Yuanpei, Peking University, Beijing 100871, China}
\affiliation[b]{Department of Astronomy, School of Physics, Peking University,
Beijing 100871, China}
\affiliation[c]{Kavli Institute for Astronomy and Astrophysics, Peking
University, Beijing 100871, China}
\affiliation[d]{School of Physics, Peking University, Beijing 100871, China}
\affiliation[e]{National Astronomical Observatories, Chinese Academy of
Sciences, Beijing 100012, China}

\emailAdd{zhzheng@stu.pku.edu.cn}
\emailAdd{zwang@pku.edu.cn}
\emailAdd{deng\_le0@stu.pku.edu.cn}
\emailAdd{ydong@pku.edu.cn}
\emailAdd{lshao@pku.edu.cn}

\abstract{Despite the large uncertainties in the equation of state for neutron
stars (NSs), a tight universal ``Love-Q'' relation exists between their
dimensionless tidal deformability, $\Lambda$, and the dimensionless quadrupole
moment, $Q$.  However, this relation has not yet been directly measured through
observations.  Gravitational waves (GWs) emitted from binary NS (BNS)
coalescences provide an avenue for such a measurement.  In this study, we adopt
a hierarchical Bayesian framework and combine multiple simulated GW events to
measure the Love-Q relation.  We simulate 1000 GW sources and select 20 events
with the highest signal-to-noise ratios and NS spins for the analysis.  By
inspecting four parameterization models of the Love-Q relation, we observe
strong correlations between the model parameters.  We verify that a linear
relation between $\ln\Lambda$ and $\ln Q$ is practically sufficient to describe
the Love-Q relation with the  precision expected from next-generation GW
detectors.  Furthermore, we utilize the inferred Love-Q relation to test
modified gravity. Taking the dynamical Chern-Simons gravity as an example, our
results suggest that the characteristic length can be constrained to $10\,
\mathrm{km}$ or less with future GW observations. 
}

\begin{document}
\maketitle
\flushbottom

%=============================
\section{Introduction}
\label{sec:introducion}
%=============================

Thanks to their extreme densities and strong gravitational fields, neutron stars
(NSs) serve as natural laboratories for studying nuclear and gravitational
physics (see e.g., Ref.~\cite{Shao:2022koz}). Inferred quantities from
electromagnetic observations, such as the observed maximum 
mass~\cite{Ozel:2010bz, Hebeler:2013nza, Antoniadis:2013pzd} and mass-radius 
relation~\cite{Lattimer:2006xb, Steiner:2010fz, Ozel:2010fw, Ozel_2013,
Guver:2013xa}, allow one to probe the properties of nuclear matter at densities
exceeding the nuclear saturation density. Additionally, the observation of
GW170817 has opened up a new window for investigating NS properties using
gravitational waves (GWs) emitted from binary NS (BNS)
coalescences~\cite{LIGOScientific:2017vwq, LIGOScientific:2018cki, 
LIGOScientific:2018hze}. Some of the NS properties, including the tidal
deformability and the spin-induced quadrupole moment, contribute to the GW
emission ~\cite{Poisson:1997ha, Vines:2011ud, Favata:2013rwa, Wade:2014vqa,
Samajdar:2019ulq, Abac:2023ujg} and therefore could be measured from GW
observations~\cite{Harry:2018hke, Baiotti:2019sew, Chatziioannou:2020pqz,
Agathos:2015uaa, Krishnendu:2017shb, Krishnendu:2019tjp, Gao:2021uus}. 

In a BNS system, each NS is deformed due to the gravitational field of its
companion, leading to an induced mass quadrupole moment~\cite{Hinderer:2007mb,
Damour:2009vw}. The effect is characterized by the tidal deformability
$\Lambda=2k_2/(3C^5)$, where $k_2$ is the tidal Love number and $C$ is the NS
compactness~\cite{Flanagan:2007ix}. Also, a rotating NS experiences another
deformation due to its spin, which induces the so-called  spin-induced
quadrupole moment, $\mathcal{Q}=-Q\chi^2 m^3$, where $m$ is the mass of the NS,
$\chi$ and $Q$ are the dimensionless spin and quadrupole moment
respectively~\cite{Hartle:1968, Laarakkers:1997hb}.  These properties can
provide an insight into the internal structure of NSs and bring opportunities to
test strong-field gravity~\cite{Akmal:1998cf, Demorest:2010bx, Ozel:2016oaf,
NANOGrav:2019jur, Li:2020wbw, Hu:2021tyw, Dong:2023vxv}. 

The internal structure of NSs depends on both the underlying gravity theory and
the equation of state (EOS); the latter describes the relation between pressure
and density of the NS matter. Despite starting to place constraints on the EOS,
current observations are in general not accurate enough to distinguish between
various EOS candidates~\cite{Lattimer:2006xb, Steiner:2010fz, Ozel:2010fw,
Hebeler:2013nza, Ozel_2013}.  This means that when it comes to testing
strong-field gravity, one will encounter a degeneracy between the EOS and the
underlying gravity theory~\cite{Yagi:2013bca, Yagi:2013awa, Shao:2017gwu,
Shao:2019gjj, Silva:2020acr, Shao:2022koz}.  One way to break this degeneracy is
to find universal relations among NS properties.  Yagi and
Yunes~\cite{Yagi:2013bca, Yagi:2013awa} found such a universal relation between 
$I$ (moment of inertia), $\Lambda$ (tidal Love number) and $Q$ (quadrupole
moment), assuming the validity of general relativity (GR).  On the one hand,
this ``I-Love-Q'' relation is insensitive to EOS uncertainties with variations
of about $1\%$ or less for different EOSs.  On the other hand, the relation can
deviate from the GR prediction in modified gravity theories~\cite{Yagi_2017,
Gupta:2017vsl, Yunes:2025xwp}, enabling an EOS-insensitive test of gravity.
Among the I-Love-Q trio, the Love-Q relation can be probed with GW observations,
since both $\Lambda$ and $Q$ affect the GWs emitted by BNS systems.  If the
Love-Q relation in GR is adopted as a prior in the waveform model, the number of
independent parameters can be reduced, which helps to better estimate the spin
parameters of BNSs~\cite{Yagi:2013bca, LIGOScientific:2018cki,
LIGOScientific:2018hze, LIGOScientific:2020aai}. The existence of such universal
relations, which now include many others~\cite{Lau:2009bu, Yagi:2013sva,
Maselli:2013mva, Pani:2015nua, Yagi:2016qmr, Gao:2023mwu, Hu:2025gab}, may also
imply a ``no-hair-theorem-like'' behavior for NSs, bringing us a new insight
into fundamental physics. 

Future next-generation (XG) ground-based GW detectors, including the Cosmic
Explorer (CE)~\cite{Reitze:2019iox, Reitze:2019dyk} and the Einstein Telescope
(ET)~\cite{Punturo:2010zz, Hild:2010id, Sathyaprakash:2012jk, ET:2025xjr}, are
expected to detect many more GW signals, up to about $10^5$--$10^6$ events per
year for BNS coalescences~\cite{LIGOScientific:2017zlf, Sathyaprakash:2019yqt, Kalogera:2021bya,
Samajdar:2021egv}, thanks to their increased sensitivity and lower cutoff
frequencies. These high-precision GW observations allow us to treat $\Lambda$
and $Q$ as independent parameters in the waveform model, to be measured directly
from GWs. This enables further constraints on the Love-Q relation from an
observational perspective.  Samajdar and Dietrich~\cite{Samajdar:2020xrd} have
first performed an analysis discussing the prospects of constraining Love-Q 
relation with GW observations, where a weighted linear regression was performed.
While the results are very valuable, as we will show below, such a treatment
might miss possible degeneracy and non-Gaussianity in the posteriors of
$\Lambda$ and $Q$. 

For the first time, our study adopts the hierarchical Bayesian framework to
infer the Love-Q relation with XG GW observations.  This framework has been
successfully applied in population studies of compact binary coalescences, and
in probing the EOS of NSs~\cite{Mandel:2009nx, Mandel:2009pc, Adams:2012qw,
Lackey:2014fwa, Mandel:2018mve, Golomb:2021tll, KAGRA:2021duu, Wang:2024xon}. 
Regarding the fitting parameters in the Love-Q relation as hyperparameters, the
hierarchical Bayesian framework separates the inferences of these
hyperparameters and the single-event parameters into two layers to avoid a
direct, high-dimensional inference for all unknown parameters. It significantly
reduces the computational cost in combining information from multiple events.
Also, the construction of the quasi-likelihood function in this framework
incorporates the full shape of the posterior in single-event inference beyond
Gaussianity, thus utilizes the information contained therein in a more
comprehensive way.  

In our implementation, we simulate 1000 GW events based on population models of
NSs~\cite{Fishbach:2018edt, Farrow:2019xnc, Samajdar:2020xrd} and select the 20
loudest events for analysis.  We find that the primary information for
constraining the Love-Q relation comes from the 10 loudest GW events, consistent
with results found in previous studies~\cite{Lackey:2014fwa}. In the pioneering
work of Samajdar and Dietrich~\cite{Samajdar:2020xrd}, a linear relation between
$\ln\Lambda$ and $\ln Q$ is adopted. By further considering four polynomial
models from linear to quartic terms in fitting the relation, we quantitatively
show that the linear relation is accurate enough when constraining the Love-Q
relation with GWs.  Additionally, we apply the inferred Love-Q relation in
gravity tests.  Taking the dynamical Chern-Simons (dCS) gravity as an example,
we find that the characteristic length $\xi_{\rm CS}^{1/4}$---with $\xi_{\rm
CS}$ the theory parameter---can be limited to $\lesssim 10\,{\rm km}$ with
future GW observations. 

This paper is organized as follows. In section~\ref{sec:framework} we construct 
the hierarchical Bayesian framework and derive the posterior of the
hyperparameters.  The simulation procedure is explained in
section~\ref{sec:simulation}.  We present the results of our inference and
discuss the differences between different Love-Q parameterization models in
section~\ref{sec:results}.  We compare our inference results with the
predictions in the dCS gravity in section~\ref{sec:dCS}. Finally, we conclude in
section~\ref{sec:conclusion}.

%=============================
\section{Hierarchical Bayesian Inference of Love-Q Relation}
\label{sec:framework}
%=============================

%=============================
\subsection{Polynomial Models of Love-Q Relation} 
\label{subsec:framework_parameterization}
%=============================

\begin{figure}[tbp]
\centering
\includegraphics[width=0.8\textwidth]{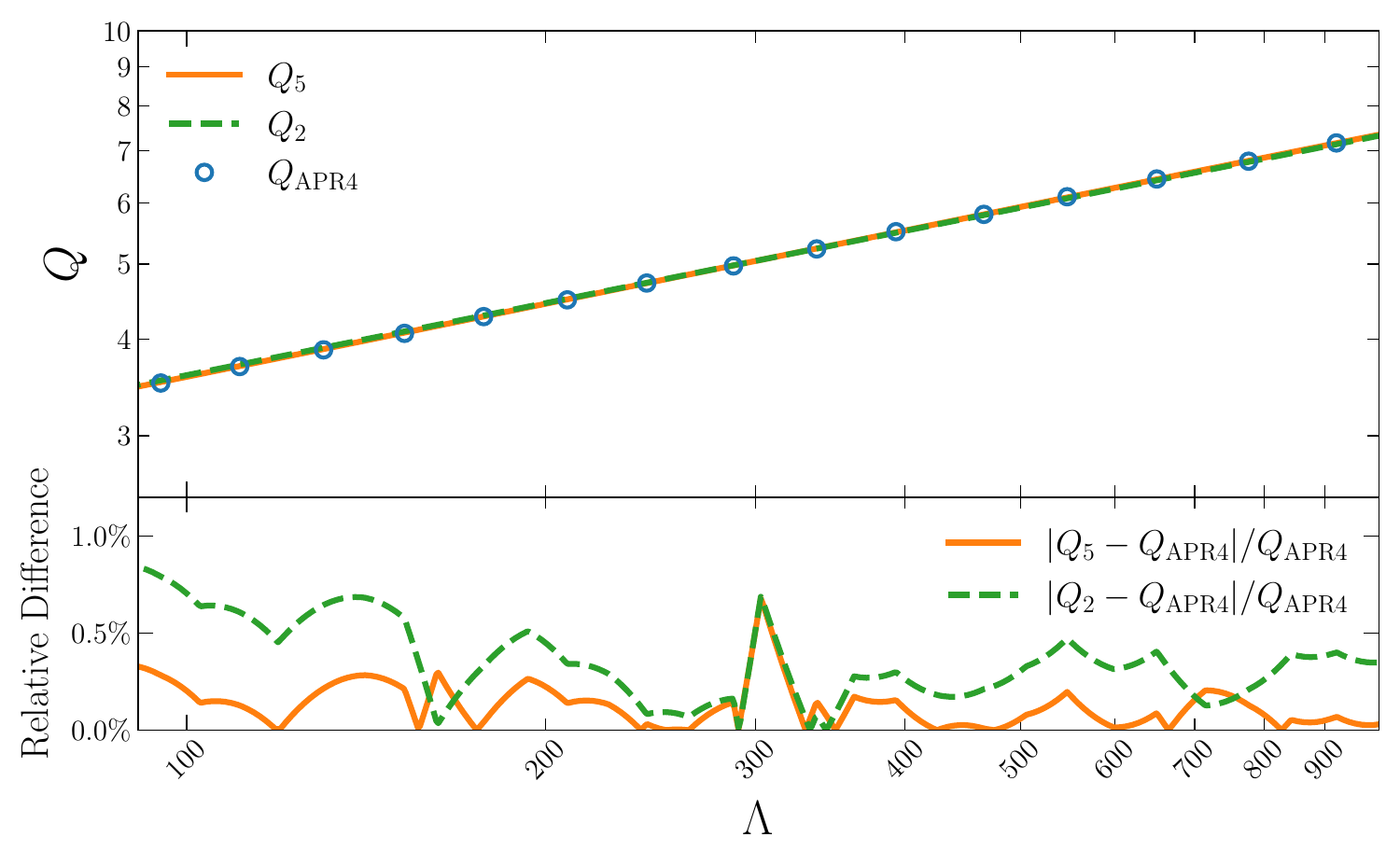}
% Here is how to import EPS art
\caption{Illustration of the Love-Q relation. In the upper panel, the orange
solid line indicates the original Yagi-Yunes relation~\eqref{5-d_Love_Q_eq},
while the green dashed line represents our fitting with the linear
model~\eqref{2-d_Love_Q_eq}. We also show the Love-Q relation for the APR4 EOS
as reference with blue circles.  In the lower panel, we show the absolute
relative differences between the two models (denoted as $Q_5$ and $Q_2$) and the
APR4 EOS (denoted as $Q_{\rm APR4}$). }
\label{relative_difference}
\end{figure}

Yagi and Yunes~\cite{Yagi:2013bca, Yagi:2013awa, Yagi_2017} fit the Love-Q
relation with a quartic polynomial model as,
%--
\begin{equation}
\label{5-d_Love_Q_eq}
    \ln Q_{5}=a_5 + b_5 \ln \Lambda + c_5 \ln^2\Lambda + d_5 \ln^3\Lambda + e_5
    \ln^4 \Lambda\,,
\end{equation}
%--
where the dimensionless fitting coefficients are $a_5=0.1940$, $b_5=0.09163$,
$c_5=0.04812$, $d_5=-4.283\times 10^{-3}$ and $e_5=1.245\times
10^{-4}$~\cite{Yagi_2017}, and the lower subscript ``$5$'' indicates five model
parameters.  They found that such a relation applies to most of the EOSs with 
relative deviation less than $1\%$ in GR. When constraining the Love-Q relation
with GWs, \citet{Samajdar:2020xrd} adopted a linear model,
%--
\begin{equation}
\label{2-d_Love_Q_eq}
    \ln Q_{2} = a_2 + b_2 \ln \Lambda\,,
\end{equation}
%--
where only two parameters, $a_2$ and $b_2$, are involved.  In
figure~\ref{relative_difference}, we plot the Yagi-Yunes
relation~(\ref{5-d_Love_Q_eq}) along with its linear fit~(\ref{2-d_Love_Q_eq}),
which is obtained by a least squares regression performed on 1000 points
uniformly picked in the logarithmic space from the Yagi-Yunes relation ($a_2=-0.1457$
and $b_2=0.3094$).  As an example of a typical EOS, we plot the
Love-Q relation calculated assuming the APR4 EOS~\cite{PhysRevC.58.1804}, a soft
EOS consistent with the observations of GW170817~\cite{LIGOScientific:2017vwq,
LIGOScientific:2018cki, LIGOScientific:2018hze}. In this case, the relative
differences in $Q$ between these two models and the APR4 EOS are smaller
than $1\%$. 

%---------------------------------------------------------------------
%---------------------------------------------------------------------

%=============================
\subsection{Hierarchical Bayesian Inference}
\label{subsec:framework_principles}
%=============================

The coefficients in Eq.~\eqref{5-d_Love_Q_eq} and Eq.~\eqref{2-d_Love_Q_eq} do
not directly contribute to the GW waveform. Instead, they determine a relation
between the waveform parameters $\Lambda$ and $Q$, denoted as
$Q=f(\Lambda;\bm{H})$, where $\bm{H}$ represents the coefficients, $\bm{H} =
\{a_2, b_2\}$ or $\bm{H} = \{a_5, b_5, c_5, d_5, e_5\}$. When the universal
relation is exact, this leads to a $\delta$-function-type prior between GW
parameters $\Lambda$ and $Q$
%--
\begin{equation}
\label{delta function prior}
\pi(Q|\Lambda,\bm{H}) = \delta\big(Q-f(\Lambda;\bm{H})\big)\,.
\end{equation}
%--
Intuitively, one can measure $\Lambda$ and $Q$ from BNS GW events, and then fit
the relation with the measurements. This procedure can be implemented within the
hierarchical Bayesian framework, which is a powerful formalism in studying population
properties of GW events beyond individual observations~\cite{Thrane_2019}. 

In the population inference scenario, the population properties are
characterized by a set of hyperparameters, such as the power index of the mass
distribution. These hyperparameters do not enter the waveform directly, while
can be inferred from a collection of measurements of single-event parameters.
The EOS parameters, which determine the $\Lambda$-$m$ and $Q$-$m$ relations,
can also be regarded as hyperparameters and analyzed in the hierarchical Bayesian framework~\cite{Mandel:2009nx, Mandel:2009pc,
Adams:2012qw, Lackey:2014fwa, Mandel:2018mve, Golomb:2021tll, KAGRA:2021duu,
Wang:2024xon}. In this work, we adopt this framework to the inference of the
Love-Q relation.

Below we briefly introduce the hierarchical Bayesian framework and describe the
customization in inferring the Love-Q relation.  The hierarchical Bayesian
framework aims to find the posterior distribution of the hyperparameters
$p(\bm{H}|D)$, given the catalog-level data $D=\{d_1,\cdots,d_n\}$ consisting of
$n$ individual GW events.  Since both the hyperparameters $\bm{H}$ and the
single-event parameters $\{\bm{\theta}_1,\cdots,\bm{\theta}_n\}$ are unknown, we
write the Bayes' theorem as 
%--
\begin{equation}
\label{bayes2}
p(\bm{H},\bm{\theta}_1, \cdots,\bm{\theta}_n|D) =\frac{p(D|\bm{H},
\bm{\theta}_1, \cdots,\bm{\theta}_n)\pi(\bm{H}, \bm{\theta}_1,
\cdots,\bm{\theta}_n)}{p(D)}\,,
\end{equation}
%--
where $\pi(\bm{H},\bm{\theta}_1,\cdots,\bm{\theta}_n)$ is the prior,
$p(D|\bm{H},\bm{\theta}_1,\cdots,\bm{\theta}_n)$ is the likelihood function, and
$p(D)$ denotes the evidence. Then, $p(\bm{H}|D)$ can be obtained by 
marginalizing over the single-event parameters
%--
\begin{equation}
\label{bayes1}
p(\bm{H}|D) = \int p(\bm{H},\bm{\theta}_1, \cdots, \bm{\theta}_n|D) \text{d}
\bm{\theta}_1\cdots\text{d} \bm{\theta}_n\,.
\end{equation}
%--

Assuming that the $n$ events are independent, the prior can be decomposed into
parts of hyperparameters and that of single-event parameters,
%--
\begin{equation}
\label{bayes3}
\pi(\bm{H}, \bm{\theta}_1, \cdots,\bm{\theta}_n) = \pi(\bm{H}) \prod_{i=1}^n
\pi(\bm{\theta}_i|\bm{H})\,.
\end{equation}
%--
We treat the tidal deformabilities, $\bm{\Lambda}_i=\{\Lambda_{1i},\Lambda_{2i}\}$, 
and quadrupole moments, $\bm{Q}_i=\{Q_ {1i},Q_{2i}\}$, of the two NSs of the $i\text{-th}$ event 
as independent parameters in $\bm{\theta}_i$ and select flat priors for
them. The impacts of priors are relatively small, since only the loudest events with high SNRs are analyzed. Note that the Love-Q
relation is insensitive to EOS, the priors of mass parameters, $m_1$ and $m_2$,
are chosen to be independent of $\bm{\Lambda}_i, \bm{Q}_i$ and $\bm{H}$. Other
parameters in $\bm{\theta}_i$ are also assumed to be independent.  In this way,
the conditional prior of $\bm{\theta}_i$ can be further decomposed as
%--
\begin{equation}
\label{prior}
\pi(\bm{\theta}_i|\bm{H}) = \pi(\bm{\Lambda}_i|\bm{H}) \,
\pi(\bm{Q}_i|\bm{\Lambda}_i,\bm{H}) \, \pi(\bm{\xi}_i)\,,
\end{equation}
%--
where $\bm{\xi}_i$---the so-called nuisance parameters hereafter---denotes the
other parameters in $\bm{\theta}_i$ except for $\bm{\Lambda}_i$ and $\bm{Q}_i$.
Analogously to the prior, the catalog likelihood can be factorized into the
product of single-event likelihoods
%--
\begin{equation}
    p(D|\bm{H},\bm{\theta}_1, \cdots,\bm{\theta}_n) = \prod_{i=1}^{n}
    p(d_i|\bm{H},\bm{\theta}_i)=\prod_{i=1}^{n}
    p(d_i|\bm{\theta}_i)\,,\label{eq:catalog_likelihood}
\end{equation}
%--
where the second equality comes from the fact that the hyperparameters do not
enter the waveform.

Given the specific form of the prior and the likelihood, the marginalized
posterior distribution \eqref{bayes1} becomes
%--
\begin{equation}
\label{hierarchical bayes}
\begin{aligned}
p(\bm{H}|D) &= \frac{1}{p(D)}\pi(\bm{H}) \int \text{d}\bm{\theta}_1
\cdots\text{d} \bm{\theta}_n \prod_{i=1}^n \big[\pi(\bm{\Lambda}_i|
\bm{H}) \, \pi(\bm{Q}_i| \bm{\Lambda}_i,
\bm{H}) \,\pi(\bm{\xi}_i) \,p(d_i|\bm{\theta}_i)\big] \\
&=\frac{1}{p(D)} \pi(\bm{H}) \prod_{i=1}^n \int \text{d}\bm{\Lambda}_i
\text{d}\bm{Q}_i \,\pi (\bm{\Lambda}_i|\bm{H}) \, \delta\big(\bm{Q}_i -
\bm{f}(\bm{\Lambda}_i; \bm{H})\big) \int \text{d} \bm{\xi}_i \,
\pi(\bm{\xi}_i)p(d_i|\bm{\theta}_i)\\
&=\frac{1}{p(D)} \pi(\bm{H}) \prod_{i=1}^n \int \text{d}\bm{\Lambda}_i \,
\pi(\bm{\Lambda}_i| \bm{H})L_i\big( \bm{\Lambda}_i,
\bm{f}(\bm{\Lambda}_i;\bm{H})\big)\,.
\end{aligned}
\end{equation}
%--
In the second line, the bold font $\bm{f}(\bm{\Lambda}_i;\bm{H})$ is used to
denote the vector form of the Love-Q relation applied to the two NSs in a BNS
system. In the third line, $L_i(\bm{\Lambda}_i,\bm{Q}_i)$, called the
quasi-likelihood, is defined as the integral over the nuisance parameters
%--
\begin{equation}
\label{quasi-likelihood}
    L_i(\bm{\Lambda}_i,\bm{Q}_i) =\int \text{d}\bm{\xi}_i \,
    \pi(\bm{\xi}_i)p(d_i|\bm{\theta}_i)\,.
\end{equation}
%--

The quasi-likelihood of each event can be computed independently of $\bm{H}$.
For the $i\text{-th}$ event, we write down the Bayes' theorem
%--
\begin{equation}
\label{single bayes}
    p(\bm{\theta}_i|d_i, \varnothing)\propto
    \pi(\bm{\theta}_i|\varnothing)p(d_i|\bm{\theta}_i)\,,
\end{equation}
%--
where $\pi(\bm{\theta}_i|\varnothing)$ denotes an auxiliary prior independent of
$\bm{H}$.  The explicit form of the single-event likelihood,
$p(d_i|\bm{\theta}_i)$, is given by assuming stationary and Gaussian
noise~\cite{Finn:1992wt}
 %--
\begin{equation}
p(d_i|\bm{\theta}_i)\propto \mathrm{e}^{ -\frac{1}{2} \langle
d_i-h(\bm{\theta}_i),d_i-h(\bm{\theta}_i)\rangle}\,,
\end{equation}
%--
with the data $d_i$ and the waveform model $h(\bm{\theta}_i)$.  The inner
product of $u(t)$ and $v(t)$, $\langle u, v\rangle$, is defined as
%--
\begin{equation}
    \langle u, v\rangle:= 2\int_{-\infty}^{\infty} \frac{\tilde{u}(f)
    \tilde{v}^{*}(f)}{S_n(|f|)} \text{d}f\,,
\end{equation}
%--
where $\tilde{u}(f)$ and $\tilde{v}(f)$ are the Fourier transforms of $u(t)$ and
$v(t)$, and $S_n(f)$ is the power spectrum density (PSD) of the noise.

Rearranging terms of Eq.~\eqref{single bayes} and substituting it into
Eq.~\eqref{quasi-likelihood}, one finds that
%--
\begin{equation}
\label{quasi-posterior}
\begin{aligned}
    L_i(\bm{\Lambda}_i,\bm{Q}_i) = \int \text{d}\bm{\xi}_i \,
    \pi(\bm{\xi}_i)p(d_i|\bm{\theta}_i) \propto \int \text{d}\bm{\xi}_i
    \frac{\pi(\bm{\xi}_i)}{\pi(\bm{\theta}_i
    |\varnothing)}p(\bm{\theta}_i|d_i,\varnothing)\,.
\end{aligned}  
\end{equation}
%--
If we further choose $\pi(\bm{\theta}_i|\varnothing) \propto\pi(\bm{\xi}_i)$, 
i.e., a flat prior for $\bm{\Lambda}_i$ and $\bm{Q}_i$,
Eq.~\eqref{quasi-posterior} can be further simplified as 
%--
\begin{equation}
\label{quasi-marginalized}
\begin{aligned}
    L_i(\bm{\Lambda}_i,\bm{Q}_i) \propto \int \text{d} \bm{\xi}_i \,
    p(\bm{\theta}_i|d_i, \varnothing)\propto p(\bm{\Lambda}_i,
    \bm{Q}_i|d_i,\varnothing)\,.
\end{aligned}  
\end{equation}
%--
This means that the quasi-likelihood is proportional to the marginalized
posterior of the auxiliary single-event inference with a flat prior on
$\bm{\Lambda}_i$ and $\bm{Q}_i$. 

The hierarchical Bayesian framework, as its name implies, introduces two levels
of inferences. The first level consists of single-event Bayesian inferences
based on Eq.~\eqref{single bayes}, where the quasi-likelihoods are constructed
according to Eq.~\eqref{quasi-marginalized}. In the second level, the
quasi-likelihoods are combined to infer the hyperparameters based on
Eq.~\eqref{hierarchical bayes}.

%=============================
\section{Simulation}
\label{sec:simulation}
%=============================

%=============================
\subsection{Waveform, Population and Detectors}
\label{subsec:simulation_preliminaries}
%=============================

In our simulation, we adopt the {\sc IMRPhenomXAS\_NRTidalv3} waveform
model~\cite{Abac:2023ujg}, which includes tidal amplitude corrections as well as
spin-induced quadrupole moment terms up to 3.5\,PN with aligned spins.  The
single-event parameters $\bm{\theta}$ include the binary masses $m_1$ and $m_2$,
the dimensionless tidal deformabilities $\Lambda_1$ and $\Lambda_2$,
spin-induced quadrupole moments $Q_1$ and $Q_2$, the dimensionless spins
$\chi_1$ and $\chi_2$, the luminosity distance $D_L$, the coalescence time $t_{c}$,
the right ascension $\alpha$ and declination $\delta$, the inclination angle
$\iota$, the GW polarization angle $\psi$, and the phase of coalescence
$\phi_{c}$.

When generating the BNS events, we adopt the population model proposed by
\citet{Farrow:2019xnc}.  According to the spin magnitude, the model divides a
BNS into a recycled NS and a nonrecycled (\emph{slow}) one, for which the masses
are labeled as $m_{\mathrm{r}}$ and $m_{\mathrm{s}}$, respectively.  The
distribution of $m_{\mathrm{r}}$ has two Gaussian components while
$m_{\mathrm{s}}$ follows a uniform distribution,
%---
\begin{subequations}
\label{mass population}
\begin{equation}
    P(m_{\mathrm{r}}) = \alpha \mathcal{N}(\mu_1, \sigma_1) + (1-\alpha)
    \mathcal{N}(\mu_2, \sigma_2)\,,
\end{equation}
\begin{equation}
    P(m_{\mathrm{s}}) = \mathcal{U}(m_{\mathrm{s}}^l, m_{\mathrm{s}}^u)\,,
\end{equation}
\end{subequations}
%---
where $\alpha=0.68$, $\mu_1=1.34\, \mathrm{M}_{\odot}$, $\sigma_1=0.02\,
\mathrm{M}_ {\odot}$, $\mu_2=1.47\, \mathrm{M}_{\odot}$, $\sigma_2=0.15\,
\mathrm{M}_{\odot}$, and $m_{\mathrm{s}}^l =1.14\, \mathrm{M}_{\odot}$,
$m_{\mathrm{s}}^u =1.46\, \mathrm{M}_{\odot}$. The NS with a larger mass is
labeled as the primary star with mass $m_1$ and the other as the secondary star
with mass $m_2$.  The tidal deformability and quadrupole moments of the binary
are calculated from the stellar mass assuming the APR4 EOS with methods
described in Refs.~\cite{Yagi:2013awa, Atta:2024ckt} and the slow rotation approximation is used. 
For arbitary spin, Refs.~\cite{Doneva:2013rha, Pappas:2013naa, Yagi:2014bxa, Doneva:2014faa} have developed new
relations insensitive to certain set of EOSs with fitting coefficients depending on the spin parameter. 
Ref.~\cite{Chakrabarti:2013tca} further suggested introducing the fitting
coefficients as a function of both spin and radius, extending the
universality to various EOSs. In these cases, our framework
will still be useful with appropriate model parameterizations and modifications.

For the spin of recycled
stars $\chi_{\mathrm{r}}$, we adopt a uniform distribution
$\mathcal{U}(-0.5,0.5)$, while for $\chi_{\mathrm{s}}$ we draw from
$\mathcal{U}(-0.1,0.1)$.  Using the cosmological parameters provided by the
Planck Collaboration~\cite{Planck:2018vyg}, we simulate 1000 GW sources,
corresponding to a few years' observation from XG detectors with the observed
local merger rate, $7.6$--$250\, \mathrm{Gpc}^{-3} \,
\mathrm{yr}^{-1}$~\cite{LIGOScientific:2025pvj, LIGOScientific:2020aai}.  These
sources are distributed uniformly in source-frame time and in the co-moving
volume with distance between 15~Mpc and 150~Mpc, also uniform in sky locations
and orientations. Without loss of generality, all GW events are injected with
$t_{c}=0$.

We select a XG detector network consisting of two CE detectors and one ET 
detector, whose sensitivities are taken as CE-2~\cite{Reitze:2019iox,
Reitze:2019dyk} and ET-D~\cite{Punturo:2010zz, Hild:2010id,
Sathyaprakash:2012jk}, respectively. The two CE detectors are positioned at the
current sites of the two LIGO detectors, while the ET detector is set at the
current location of the Virgo detector with a triangular shape. 
The current locations of LIGO and Virgo are chosen as an illustrated configuration for XG detectors.

%=============================
\subsection{Implementation}
\label{subsec:simulation_implementation}
%=============================

According to the discussion in section~\ref{subsec:framework_principles}, the
inference of the Love-Q relation can be divided into two steps.  In the
auxiliary single-event inference, the variable parameters are 
%--
\begin{align}
	\bm{\theta} = \{\mathcal{M},\eta, \Lambda_1,\Lambda_2,
	Q_1,Q_2,\chi_1,\chi_2, D_L,t_{c},\alpha, \delta,\iota,\psi,\phi_{c}\} \,.
\end{align}
%--
In the parameter estimation, the priors of $\mathcal{M}$, $\eta$, $\chi_1$,
$\chi_2$, $t_{c}$ and $\phi_{c}$ are uniform, and the prior of $D_L$ is such
that the distribution is uniform in the co-moving volume. We set isotropic
priors for the angle variables $\alpha,\delta,\iota,\psi$. For tidal and
quadrupole moment parameters, we treat $\Lambda_{\mathrm{s}}$ and
$Q_{\mathrm{s}}$ of the slow binary component as nuisance parameters, since the
spin-induced quadrupole moment is poorly estimated for NSs when the spins are
slow~\cite{Yagi:2013awa}.  According to the arguments around
Eq.~\eqref{quasi-marginalized}, we choose uniform priors for $\Lambda_1$,
$\Lambda_2$, $Q_1$ and $Q_2$ in the auxiliary inference.  To calculate the
posterior, we generate samples with the {\sc Bilby}~\cite{Ashton:2018jfp}
implementation of the {\sc nessai} sampler~\cite{Skilling:2004pqw,
Skilling:2006gxv, michael_j_williams_2025_14627250, PhysRevD.103.103006,
Williams:2023ppp}.  To calculate the integral in Eq.~\eqref{hierarchical bayes},
we need the functional form of the quasi-likelihood, which is proportional to
the distribution function of the posterior $p(\bm{\Lambda}_i,
\bm{Q}_i|d_i,\varnothing)$. We follow \citet{Golomb:2021tll} and adopt the
Gaussian mixture model to estimate the density of the posterior samples.  In the
second step, the priors of the hyperparameters, $\pi(\bm{H})$, are listed in
table~\ref{prior_table}. For the conditional prior $\pi(\bm
{\Lambda}_i|\bm{H})$, we choose the uniform distribution $\mathcal{U}(10,2000)$.

We simulate 1000 GW events as described in
section~\ref{subsec:simulation_preliminaries}.  Though the hierarchical
procedure avoids a direct high-dimensional inference for $\bm{H}$ and
$\{\bm{\theta}_1,\cdots,\bm{\theta}_n\}$ simultaneously, the computational cost
still increases with the number of events $n$.  \citet{Lackey:2014fwa} found
that when constraining the EOS of NSs, several events with the highest
signal-to-noise ratios (SNRs) contribute most of the information. Also,
\citet{Yagi:2013awa} concluded that the spin-induced quadrupole moment cannot be
well measured for NSs with low spins. Considering these findings, we first draw
100 sources with the highest SNRs, then further select 20 sources with the
largest $|\chi_{\mathrm{r}}|$ among them. The SNR values of these selected events range from about 1500 to 2500.
Here we omit the selection effects 
since different $\Lambda$ and $Q$ values do not lead to a 
significant change of SNR and thus the detection probability. As we will show in
section~\ref{sec:results}, the primary information for constraining the Love-Q
relation comes from the loudest GW events. 
In practice, selecting the top 20 events is enough to capture the main
information for constraining the Love-Q relation. We present the results of
changing the number of selected events in appendix~\ref{sec: appendix number of events}.
We leave a more comprehensive study of the whole population considering selection effects for future work.

%---------------------------------------------------------------------
\begin{table}[t]
    \centering
    \sisetup{
        table-align-text-post = false, 
        separate-uncertainty = true 
    }
        \caption{The reference values and priors of the coefficients in
        different models of the Love-Q relation, with $j$ being the number of
        parameters in the polynomial model. Note that $j=5$ indicates the
        original Yagi-Yunes relation~\cite{Yagi_2017}. For $j = $ 2, 3 and 4,
        the values are the results of least squares regression performed on 1000
        points uniformly picked in the logarithmic space from the Yagi-Yunes
        relation. Priors are chosen to be uniform distributions, and the ranges
        are the same for different $j$ to ensure a meaningful comparison.
        }\label{prior_table}
    \begin{tabular}{
        l
        S[table-format=-1.4]
        S[table-format=1.5]
        S[table-format=1.3e-1]
        S[table-format=-1.3e-1]
        S[table-format=1.3e-1]
    }
        \toprule
        \multicolumn{6}{c}{Reference Values} \\
        %\cmidrule(l){2-6}
        $j$ & {$a_j$} & {$b_j$} & {$c_j$} & {$d_j$} & {$e_j$} \\
        \midrule

        5 \, &  0.1940 & 0.0916 & 4.812e-2 & -4.283e-3 & 1.245e-4 \\
        4 \, &  0.1290 & 0.1480  & 3.021e-2 & -1.817e-3 & {--}      \\
        3 \, & -0.0709 & 0.2775  & 3.220e-3 & {--}       & {--}      \\
        2 \, & -0.1457 & 0.3094  & {--}      & {--}       & {--}      \\
        
        \midrule

        \multicolumn{6}{c}{Priors} \\
        %\cmidrule(l){2-6}
        $j$ & {$a_j$} & {$b_j$} & {$c_j$} & {$d_j$} & {$e_j$} \\
        \midrule

        5 \, & {$\mathcal{U}(-5.0, 5.0)$} & {$\mathcal{U}(-1.0, 1.0)$} & {$\mathcal{U}(-0.5, 0.5)$} & {$\mathcal{U}(-0.1, 0.1)$} & {$\mathcal{U}(-0.01, 0.01)$} \\
        4 \, & {$\mathcal{U}(-5.0, 5.0)$} & {$\mathcal{U}(-1.0, 1.0)$} & {$\mathcal{U}(-0.5, 0.5)$} & {$\mathcal{U}(-0.1, 0.1)$} & {--}                         \\
        3 \, & {$\mathcal{U}(-5.0, 5.0)$} & {$\mathcal{U}(-1.0, 1.0)$} & {$\mathcal{U}(-0.5, 0.5)$} & {--}  & {--}   \\
        2 \, & {$\mathcal{U}(-5.0, 5.0)$} & {$\mathcal{U}(-1.0, 1.0)$} & {--}  & {--}   & {--}  \\
        \bottomrule
    \end{tabular}
\end{table}
%---------------------------------------------------------------------

%=============================
\section{Results and Discussions}
\label{sec:results}
%=============================

In this section, we present the inference results for different parameterization
models of the Love-Q relation. In section~\ref{subsec:results_linear_model}, we 
show the results of the linear model, Eq.~\eqref{2-d_Love_Q_eq}, consisting of
two parameters.  In section~\ref{subsec:results_quartic}, we present the results
of the quartic polynomial model in Eq.~\eqref{5-d_Love_Q_eq}, consisting of five
parameters. The polynomial models in between, i.e., the quadratic and cubic
polynomial models, are discussed in
section~\ref{subsec:results_quadratic_cubic}.
For other parameterizations beyond polynomial models, our methodology is in principle applicable to them as well.

%=============================
\subsection{The Linear Model}
\label{subsec:results_linear_model}
%=============================

%---------------------------------------------------------------------
\begin{figure}[t]
    \centering
    \includegraphics[width=0.5\linewidth]{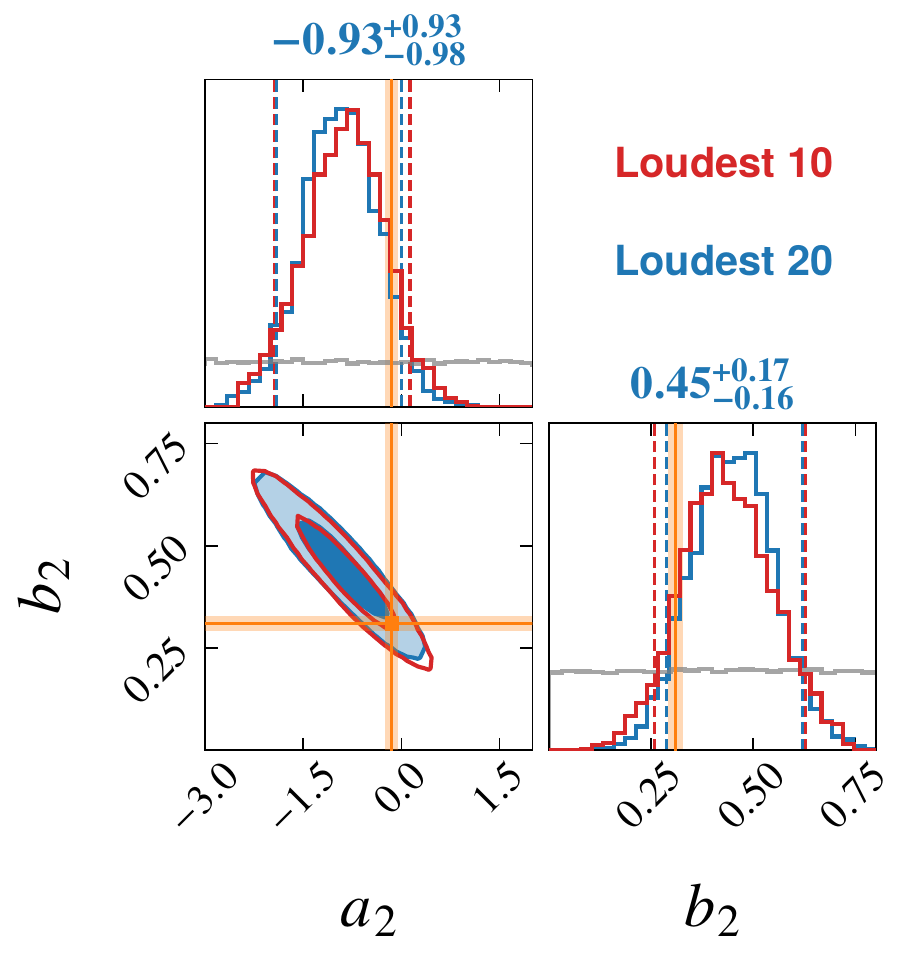}
    \caption{Posterior distributions of the hyperparameters ${\bm H} =
    \{a_2,b_2\}$ in the linear fitting model. The contours refer to 50\% and
    90\% credible regions, while the numbers above the histograms on the
    diagonal stand for the median and the central 90\% credible interval of the marginalized distribution.
    We use blue and red colors to represent the results based on the loudest 20
    and 10 events from the 1000 simulated events, respectively. 
    The orange bands represent the reference values with fitting uncertainties accounted for. The grey lines
    on the diagonal represent the priors for comparison.}
    \label{corner2-d}
\end{figure}
%---------------------------------------------------------------------

We first perform the inference for the linear model, which is also the model 
adopted in Ref.~\cite{Samajdar:2020xrd}. The posterior distribution of the 
hyperparameters $\{a_2,b_2\}$ is shown in figure~\ref{corner2-d}. Compared to
the priors in table~\ref{prior_table}, the posteriors are significantly 
narrowed. As a comparison, we also mark the fitting values of the
hyperparameters in a direct fit of the Yagi-Yunes Love-Q relation, and regard
them as ``true'' values of the linear model for reference. These values fall
within the 90\% credible region of the posterior. 
Given that the Love-Q relation is quasi-universal with relative differences of $\sim 1\%$ for various EOSs, 
the fitting uncertainties of the hyperparameters are also considered in the reference values. 
To check the robustness of our conclusions, we also repeat the inference assuming the SLy EOS; which are summarized in appendix~\ref{sec: appendix results for SLy}.
The hierarchical Bayesian
inference successfully recovers the Love-Q relation under the linear
parameterization. We also investigate how the results depend on the number of
events in the analysis. In the 20 events selected in 
section~\ref{subsec:simulation_implementation}, we further select the loudest 10
events, perform the inference again, and show the results in figure~\ref
{corner2-d}. For both the joint and marginalized distributions, the widths of
the credible regions in the 10-event inference are only slightly larger than
those in the 20-event inference. This reveals that the loudest 10 events
dominate the information in constraining the Love-Q relation, and including
quieter events will not significantly change the results.

%---------------------------------------------------------------------
\begin{figure}[t]
\centering
\includegraphics[width=\textwidth]{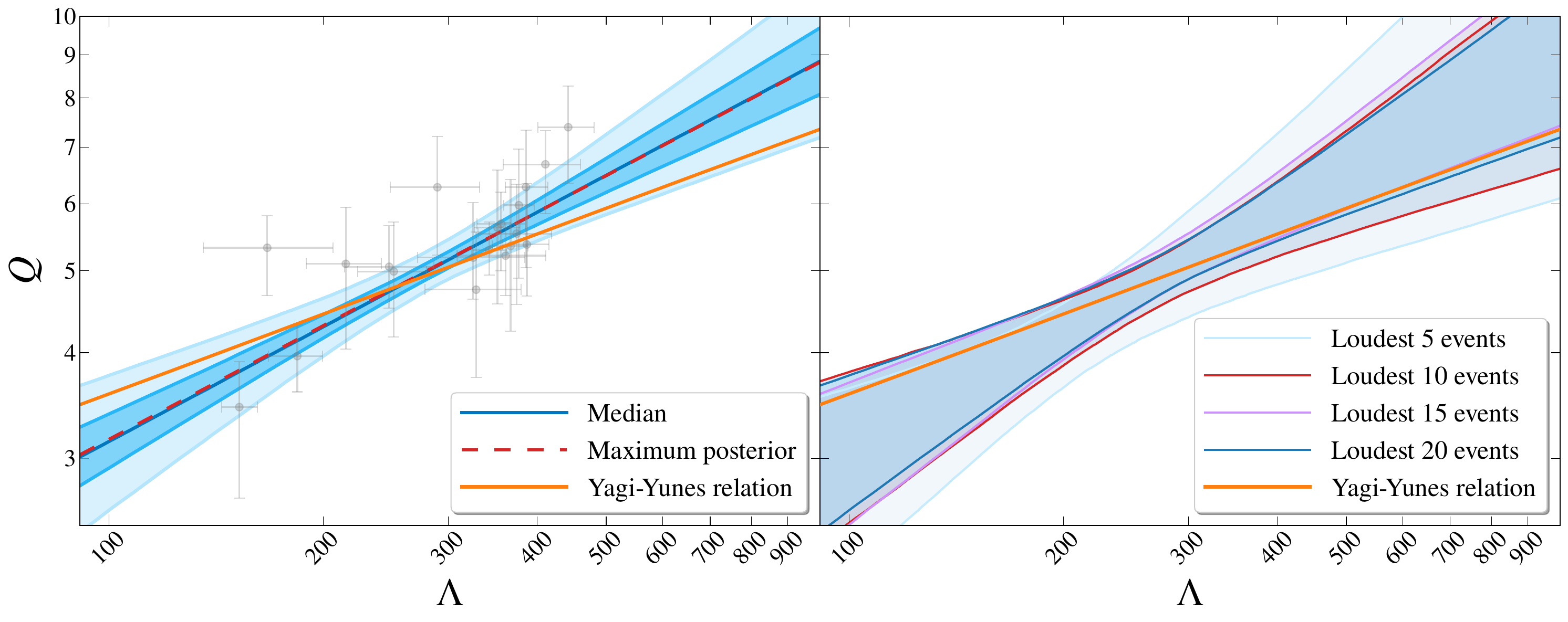}
    \caption{Recovered Love-Q relation from the posterior of the hyperparameters
    in the linear model. In the left panel, the Love-Q relation is inferred with
    the 20 loudest events from the 1000 simulated GW events. The gray points
    mark the median values of inferred $\Lambda$ and $Q$ for each event with
    $68\%$ errorbars. The blue solid line marks the median of the distribution
    of $Q$ as a function of $\Lambda$, accompanied by the $50\%$ and $90\%$
    credible intervals in shaded regions.  The red dashed line represents the
    maximum-posterior Love-Q relation. For comparison, we plot the original
    Yagi-Yunes Love-Q relation~\cite{Yagi_2017} in orange. The right panel shows
    how the $90\%$ credible region of the recovered Love-Q relation depends on
    the number of events, marked with different colors.}    
    \label{2-d_Love_Q}
\end{figure}
%---------------------------------------------------------------------

In figure~\ref{2-d_Love_Q}, we show the recovered Love-Q relation according to
the posterior samples. For each $\Lambda$, every sample in the posterior of the 
hyperparameters corresponds to a $Q$ value. For a fixed $\Lambda$ value, we find
the credible intervals of $Q$, and then vary $\Lambda$ continuously to form
credible regions. In the left panel, we show the results of the 20-event
inference. Similar to figure~\ref{corner2-d}, the Yagi-Yunes Love-Q relation is
covered by the 90\% credible region.  In the right panel, we select the loudest
5, 10, 15 and 20 events from the 20 events to test how the recovered Love-Q
relation depends on the number of events. We find that the widths of the 90\%
credible regions are almost the same for the inferences from 10, 15 and 20
events. This is consistent with the posteriors in figure~\ref{corner2-d}, and
again indicates that the loudest 10 events are almost sufficient to constrain
the Love-Q relation. Similar phenomena were also found in previous
studies~\cite{Lackey:2014fwa, Landry:2020vaw, Pang:2020ilf, Finstad:2022oni,
Bandopadhyay:2024zrr, Wang:2024xon}, where the recovered $\Lambda$-$m$ relation
is dominated by the several loudest events.

%=============================
\subsection{The Quartic Polynomial Model}
\label{subsec:results_quartic}
%=============================

%---------------------------------------------------------------------
\begin{figure}
\begin{minipage}[t]{0.49\textwidth}
\centering
\includegraphics[width=0.8\linewidth]{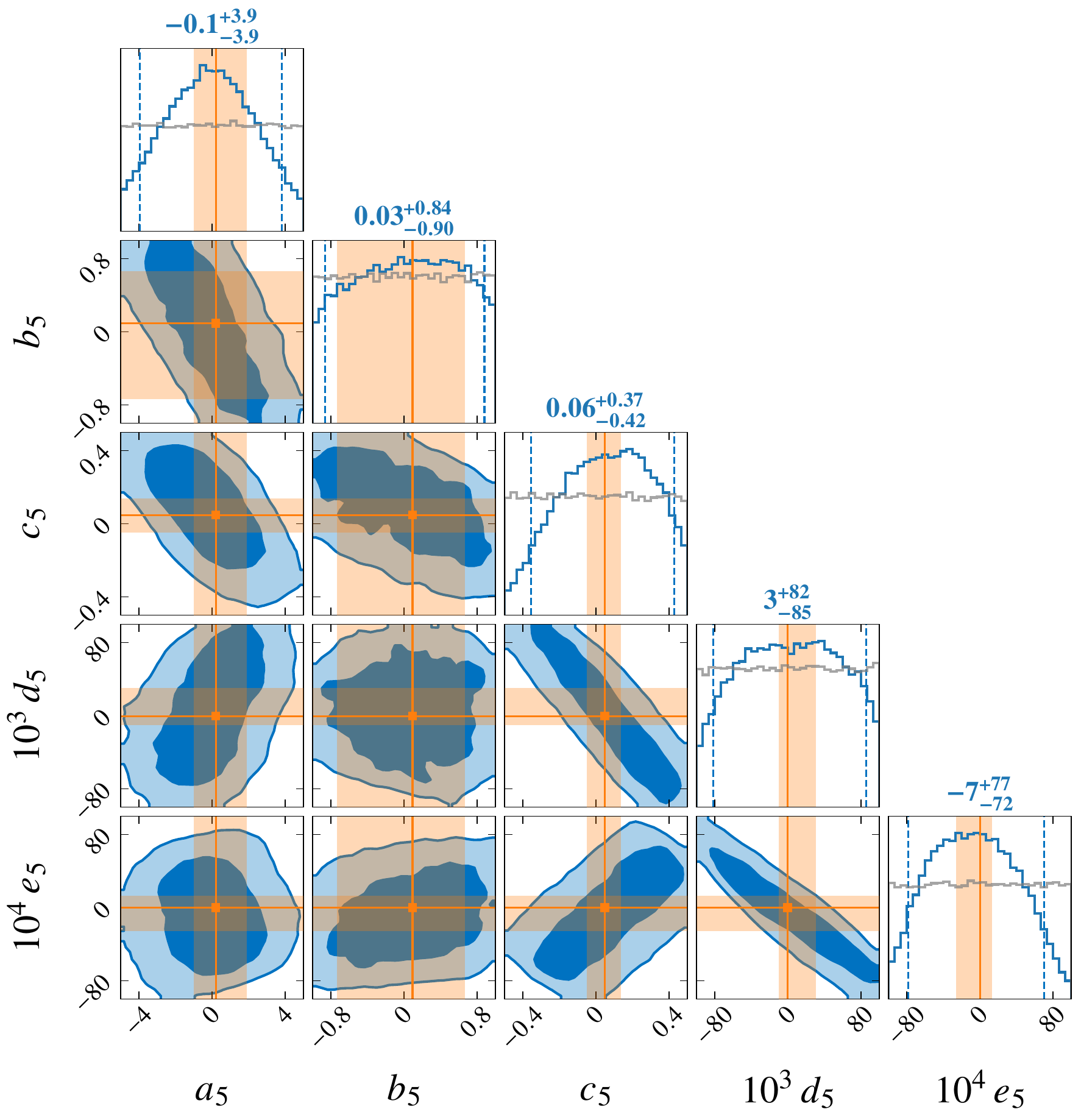}
\end{minipage}
\hfill
\begin{minipage}[t]{0.49\textwidth}
\includegraphics[width=\linewidth]{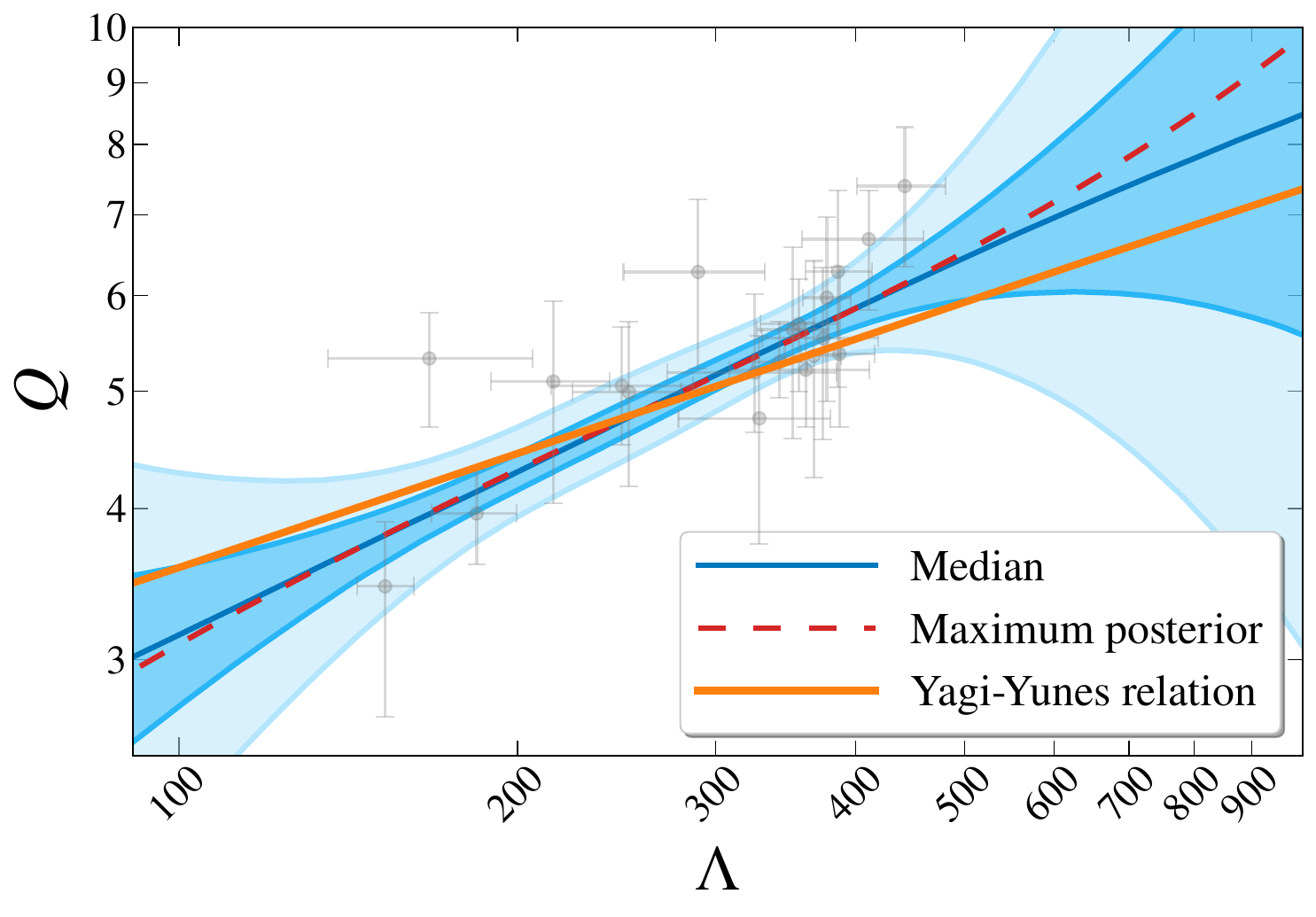}
\end{minipage}
    \caption{The posterior of the hyperparameters and the recovered Love-Q
    relation in the quartic polynomial model, where all 20 loudest events are
    included in the inference. Figure settings of two panels are respectively
    similar to those in figure~\ref{corner2-d} and the left panel of
    figure~\ref{2-d_Love_Q}. 
    } \label{5-d_Love_Q} 
\end{figure}
%---------------------------------------------------------------------

For the quartic polynomial model, the Love-Q relation is fitted with five 
parameters shown in Eq.~\eqref{5-d_Love_Q_eq}, i.e., $\{a_5, b_5, c_5, d_5,
e_5\}$.  This is  the original model proposed by Yagi and
Yunes~\cite{Yagi:2013awa}. We summarize the posterior and the recovered Love-Q
relation in figure~\ref{5-d_Love_Q}, where all 20 events are included in the
inference. In the left panel, though the true values (the values in the
Yagi-Yunes Love-Q relation~\cite{Yagi:2013awa}) are almost centered in the
distribution, the posteriors are much wider than those in the linear case,
reaching the prior boundaries. This indicates that the $\Lambda$ and $Q$
measurements from these events are not informative enough to well constrain all
the five parameters. In other words, the observation precision is not high
enough to capture the higher-order terms introduced by the additional three
parameters, $c_5, d_5$, and $e_5$. This is also reflected in the strong
correlations between the three parameters shown in the left panel of
figure~\ref{5-d_Love_Q}. In the right panel of figure~\ref{5-d_Love_Q}, the 90\%
credible regions of the recovered Love-Q relation are also wider than that in
the linear case, especially for large $\Lambda$ where the higher-order terms
become more important. For $\Lambda \sim 400$, the widths of the 90\% credible
region between the two models are similar, which is consistent with the fact
that most of the simulated events gather around this region. 

%=============================
\subsection{The Quadratic and Cubic Polynomial Models}
\label{subsec:results_quadratic_cubic}
%=============================

%---------------------------------------------------------------------
\begin{figure}[t]
    \begin{minipage}[t]{0.49\textwidth}
    \centering
        \includegraphics[width=0.8\linewidth]{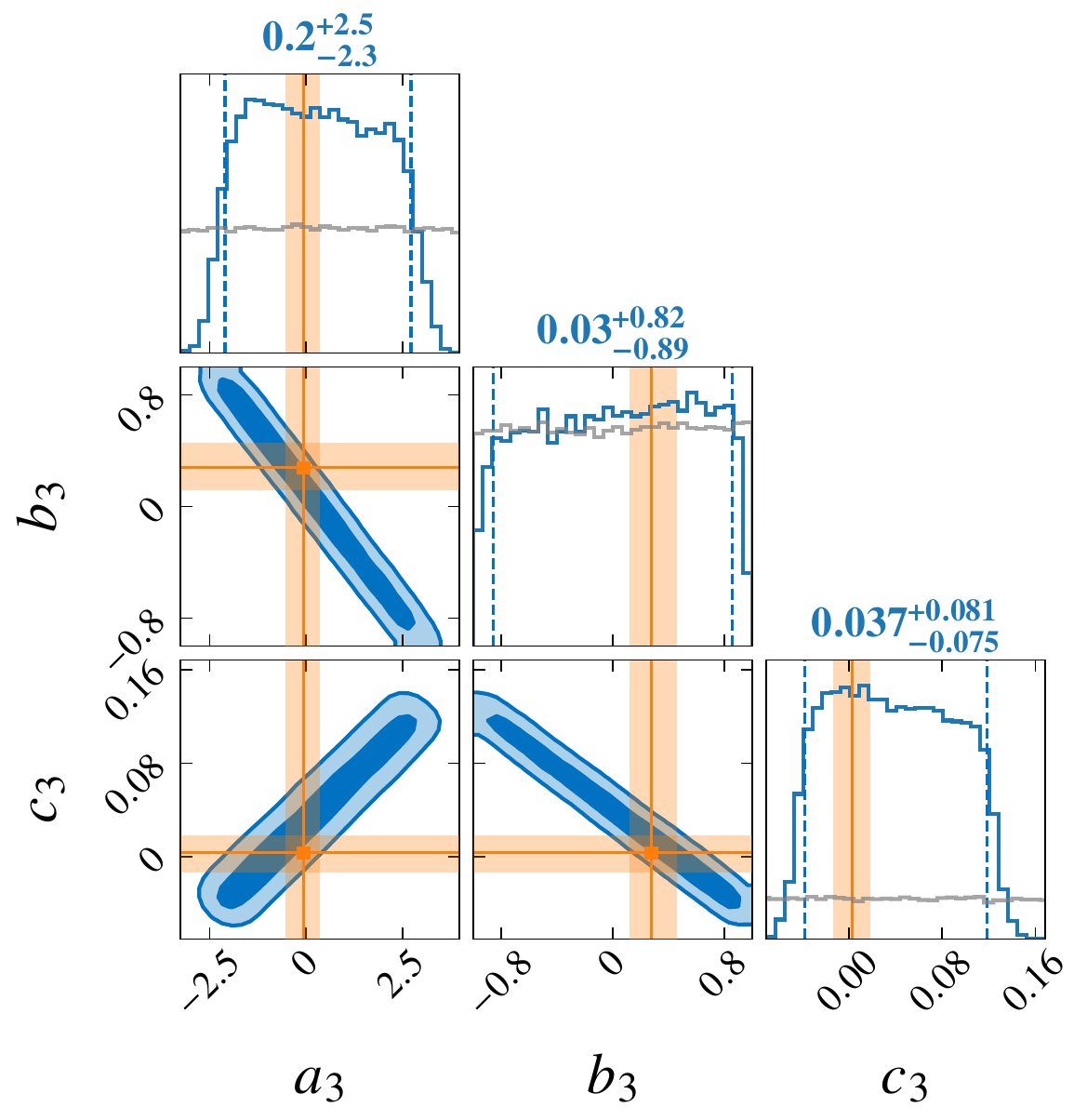}
    \end{minipage}
    \hfill
    \begin{minipage}[t]{0.49\textwidth}
        \includegraphics[width=\linewidth]{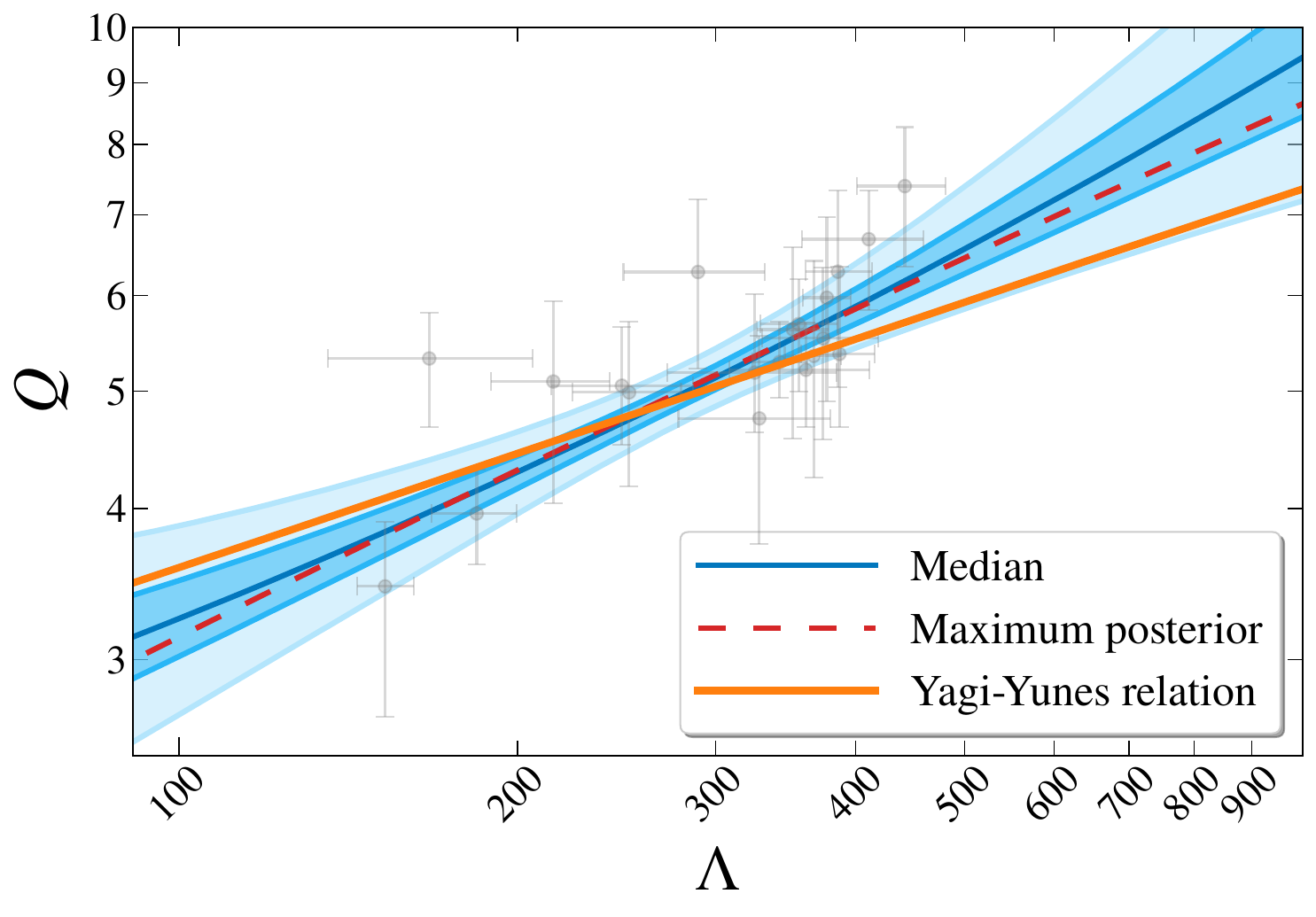}
    \end{minipage}

    \vspace{3mm}

    \begin{minipage}[t]{0.49\textwidth}
    \centering
        \includegraphics[width=0.8\linewidth]{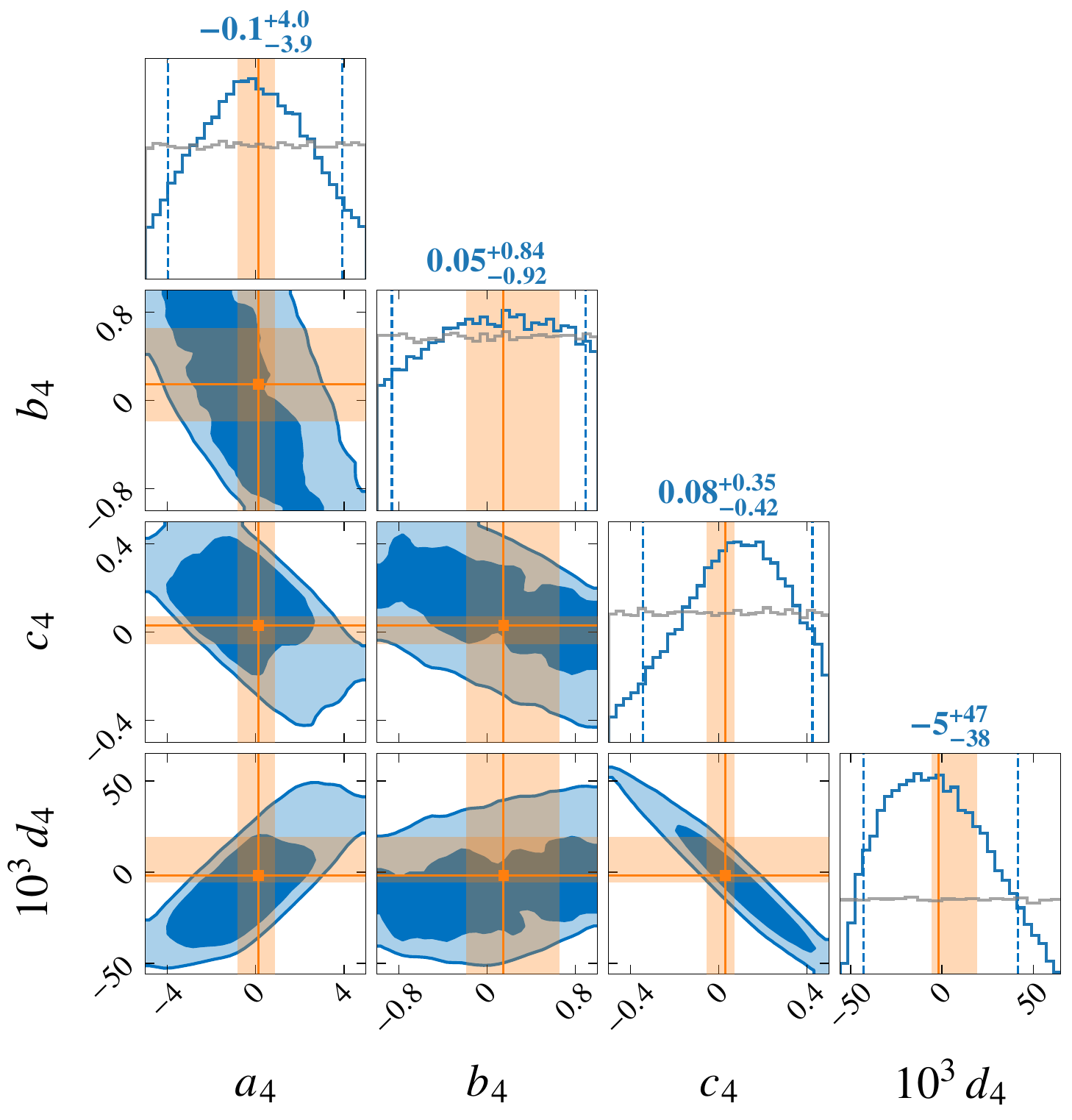}
    \end{minipage}
    \hfill
    \begin{minipage}[t]{0.49\textwidth}
        \includegraphics[width=\linewidth]{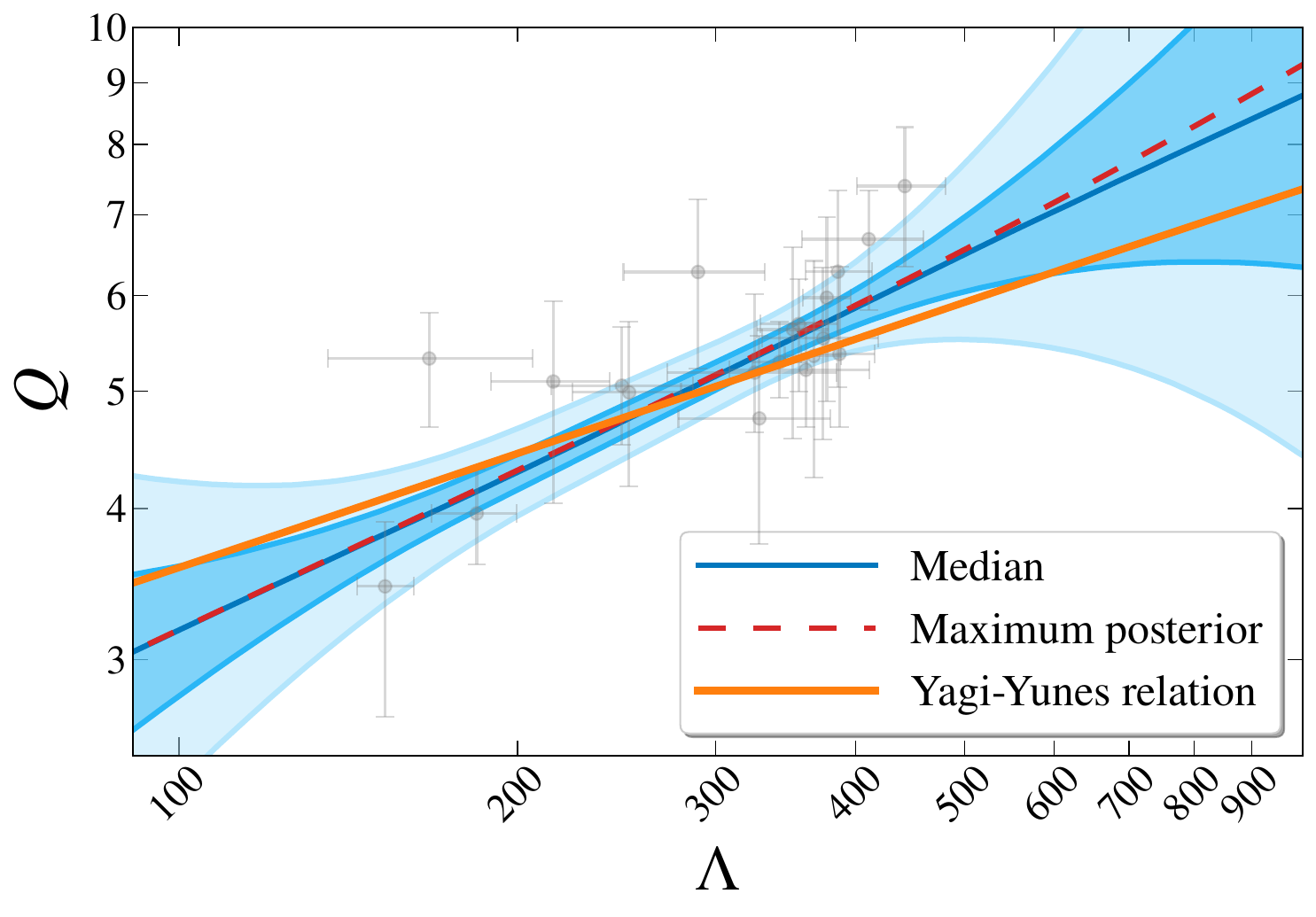}
    \end{minipage}
    \caption{Similar to figure~\ref{5-d_Love_Q}, while for the quadratic and
    cubic polynomial models. 
    }\label{3-d_4-d_Love_Q}
\end{figure}
%---------------------------------------------------------------------

As shown in the previous two subsections, the observations from 20 simulated GW 
events can well constrain the two parameters in the linear model, while cannot 
effectively constrain the five parameters in the quartic polynomial model. The
above two models were adopted in previous studies~\cite{Yagi:2013awa,
Samajdar:2020xrd}, and here we further test the models in between them and see
how the constraints change with the number of hyperparameters. We perform the
inference for the quadratic and cubic polynomial models, which contain three
parameters, $\{a_3, b_3, c_3\}$, and four parameters, $\{a_4, b_4, c_4, d_4\}$,
respectively. Similar to the linear model, the reference values of these
parameters are obtained by directly fitting the Yagi-Yunes Love-Q relation and
are listed in table~\ref{prior_table}. The priors of the hyperparameters are
kept the same as their counterparts in the quartic polynomial model, which are
also listed in table~\ref{prior_table}.

We summarize the posterior distributions of the hyperparameters in the top row
of figure~\ref{3-d_4-d_Love_Q}. In the quadratic model, strong degeneracy arises
among the three parameters. The posterior of $b_3$ is almost the same as its 
prior. For $a_3$ and $c_3$, though the posteriors are narrower than the priors, 
their marginal distributions have wide and flat plateau. In the cubic model, 
strong degeneracy  exists between $c_4$ and $d_4$, and the posterior of $b_4$ is
similar to its prior. In general, posterior with large correlations and 
prior-like marginal distributions indicates redundant parameters in the model. 
Therefore, we conclude that the linear model is accurate enough in constraining 
the Love-Q relation with XG GW observations. This is consistent with the
argument in Ref.~\cite{Samajdar:2020xrd} based on qualitative analysis, whereas
in this work we present a quantitative demonstration. 

Same as before, we plot recovered Love-Q relations in the lower panels of 
figure~\ref{3-d_4-d_Love_Q}. We find that the widths of the 90\% credible
regions around $\Lambda \sim 400$ are similar for all of four models. While for
$\Lambda$ away from this region ($\Lambda \lesssim 200$ or $\Lambda \gtrsim
500$), the widths increase with the number of parameters. This may have resulted
from the increased model complexity and the lack of data points in these
regions. 

%=============================
\section{Testing Modified Gravity: Dynamical Chern-Simons Gravity}
\label{sec:dCS}
%=============================

%---------------------------------------------------------------------
\begin{figure}[t]
    \centering
    \begin{minipage}{0.6\linewidth}
        \includegraphics[width=\linewidth]{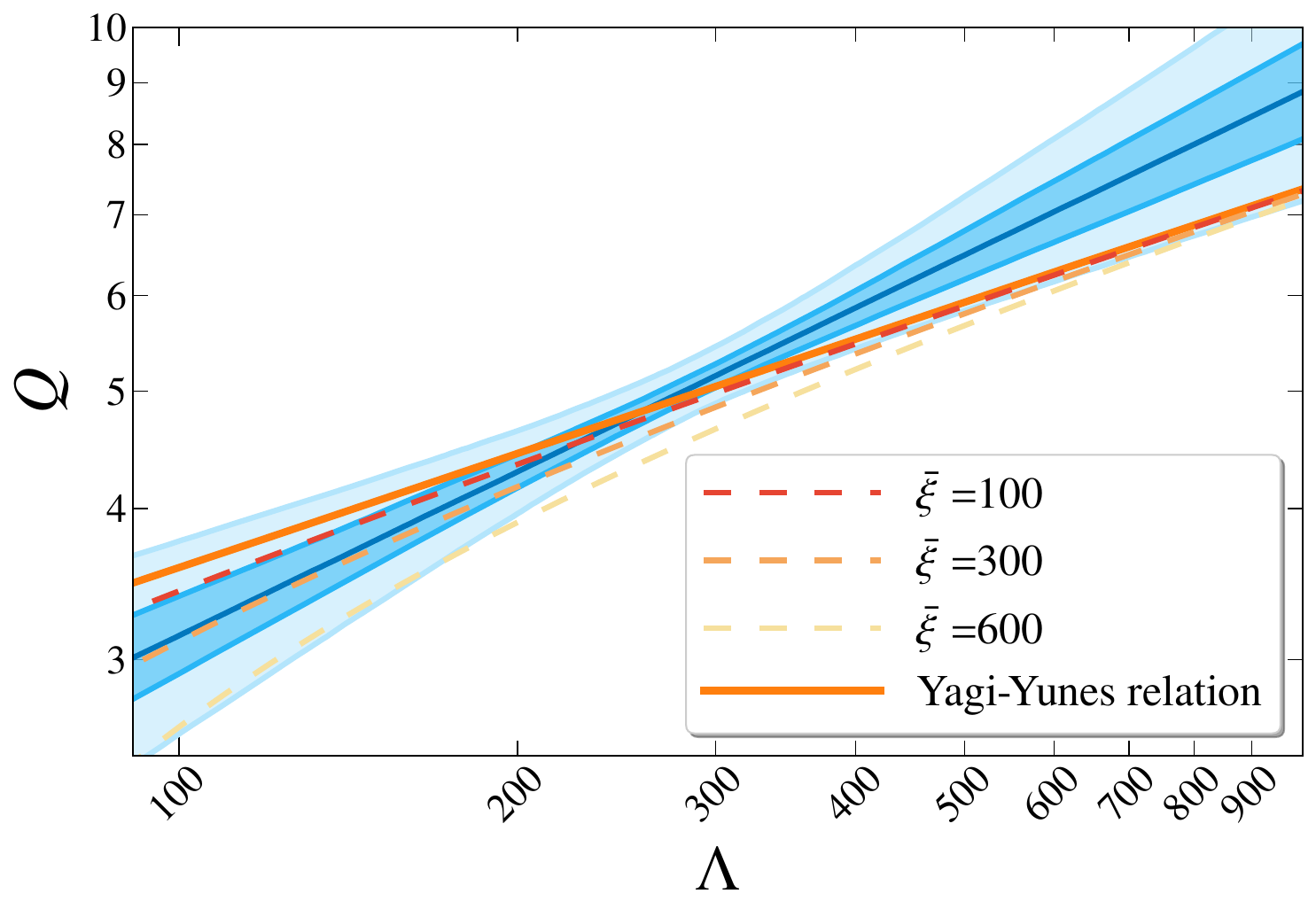}
    \end{minipage}
    \caption{Comparison of the recovered Love-Q relation in GR and that are
    predicted by the dCS gravity with coupling constant $\bar{\xi}$ fixed.
    Similar to figure~\ref{2-d_Love_Q}, the blue lines indicate the median of
    the recovered Love-Q relation, while the shaded regions represent the $50\%$
    and $90\%$ credible intervals. For the coupling constant $\bar\xi$, we take
    three values and plot the corresponding Love-Q relations.}
    \label{cs_Love_Q}
\end{figure}
%---------------------------------------------------------------------

In this section, we investigate the potential of testing gravity theories with
the inferred Love-Q relation from the GW observations. The I-Love-Q test can be 
powerful when significant difference in I-Love-Q relation between GR and
modified gravity theories exists~\cite{Shao:2022koz}, for example, in some
parity-violating theories~\cite{Yagi_2017, Yunes:2025xwp}.  Here we take the
dynamical Chern-Simons (dCS) gravity~\cite{Jackiw:2003pm, Smith:2007jm,
Alexander:2009tp} for illustration. 

The dCS gravity is well-motivated from heterotic superstring theory, loop
quantum gravity, and effective field theories of inflation~\cite{Jackiw:2003pm,
Smith:2007jm, Alexander:2009tp}.  The dCS gravity introduces parity violation
and quadratic curvature terms in the action~\cite{Alexander:2009tp,
Gupta:2017vsl},

%--
\begin{equation}
   \label{cs_action}
   S = \int \mathrm{d}^4 x \sqrt{-g}\left[ \kappa_g \mathcal{R} +
   \frac{\alpha}{4} \mathcal{\vartheta} \mathcal{R}_{\nu\mu\rho\sigma}
   {}^{*}\mathcal{R}^{\mu\nu\rho\sigma} -
   \frac{\beta}{2}\nabla_{\mu}\mathcal{\vartheta}\nabla^{\mu}\mathcal{\vartheta}
   + \mathcal{L}_{\mathrm{mat}}\right]\,,
\end{equation}
%--
where $g$ is the determinant of the metric, $\kappa_g= 1/16\pi$, $\mathcal{R}$
is the Ricci scalar, $\mathcal{R}_{\nu\mu\rho\sigma}$ and
$^{*}\mathcal{R}^{\mu\nu\rho\sigma}$ denote the Riemann tensor and its dual, 
$\mathcal{L}_{\mathrm{mat}}$ is the matter Lagrangian density, $\alpha$ and
$\beta$ are the coupling constants in the dCS gravity and the potential for the
pseudo-scalar field $\mathcal{\vartheta}$ is omitted.  $\mathcal{\vartheta}$ and
$\beta$ are taken to be dimensionless, so $\alpha$ has the dimension of length
squared, and the quantity $\xi_{\mathrm{CS}}^{1/4} \equiv [\alpha^2/
(\kappa\beta)]^{1/4}$ can be explained as the characteristic lengthscale of the
dCS gravity~\cite{Yunes:2009hc, Yagi:2012ya}.  Current Solar System observations
have constrained it to $\xi_{\mathrm{CS}}^{1/4} <\mathcal{O}(10^8
)$~km~\cite{Ali-Haimoud:2011zme, Yagi:2012ya}.

The dCS gravity predicts Love-Q relations that deviate from the one in
GR~\cite{Yagi:2013bca, Yagi:2013awa, Gupta:2017vsl}, allowing us to test the dCS
gravity with the inferred Love-Q relation.  For a Love-Q test,
\citet{Yagi:2013mbt} have obtained the dCS correction to the NS quadrupole
moment, and \citet{Yagi:2011xp} indicates that, regarding the dCS gravity as an
effective theory, the tidal deformability is the same as  in GR at leading order
in the small coupling approximation, $\zeta \equiv \xi_{\mathrm{CS}} m^2/R^6 \ll
1$. Refs.~\cite{Yagi_2017, Yagi:2013mbt, Gupta:2017vsl} have discussed the
Love-Q relation in the dCS gravity and found that the relation becomes
EOS-sensitive with $\xi_{\mathrm{CS}}$ or $\zeta$ fixed. However, with 
$\bar{\xi}\equiv \xi_{\mathrm{CS}}/m^4$ fixed, the Love-Q relation remains 
universal and the  variation with respect to EOS is of $\mathcal{O}(1\%)$.
\citet{Gupta:2017vsl} parameterized the relation between the dCS correction to
$Q$ (denoted as $Q_{\mathrm{CS}}$) and the tidal deformability $\Lambda$ as
%--
\begin{equation}
    \label{cs_Love_Q_eq}
    \ln (Q_{\mathrm{CS}}/\bar{\xi}) = a + b \ln \Lambda + c \ln^{2} \Lambda \,,
\end{equation} 
%--
where the fitting coefficients are $a=-3.443, b=-0.550$, and $c=-0.023$. 

We compare the inferred constraints of the Love-Q relation using the linear
model from figure~\ref{2-d_Love_Q} and Love-Q relations~\eqref{cs_Love_Q_eq}
with $\bar{\xi}$ fixed in the dCS gravity. From figure~\ref{cs_Love_Q} we
conclude that our Love-Q test of the dCS gravity can constrain the coupling
constant to  $\bar{\xi} \lesssim 10^{3}$. Substituting a typical mass
$m=1.4\,\mathrm{M_{\odot}}$ and a typical radius $R=10$\,km for NSs, we have
constraints for the other two coupling constants, $\xi_{\mathrm{CS}}^{1/4}
\lesssim 10$~km and $\zeta \lesssim 0.1$, which are in agreement with the
results given by Refs.~\cite{Yagi:2013bca, Yagi:2013awa}. 

%=============================
\section{Conclusion}
\label{sec:conclusion}
%=============================

In this work, we investigate the prospects of inferring the Love-Q relation of
NSs with future ground-based GW observations.  Extending the inference procedure
by \citet{Samajdar:2020xrd}, we adopt the hierarchical Bayesian framework to
effectively combine the information from multiple GW events, extending the
linear fitting model in Ref.~\cite{Samajdar:2020xrd}. The hierarchical Bayesian
framework separates the inferences into two steps, the auxiliary single-event
inference and the hyperparameter inference, which avoids a direct
high-dimensional inference for all the parameters simultaneously while still
takes into account degeneracy and non-Gaussianity in the single-event
parameters. We also discuss the impact of the event number on the constraints,
and find that the loudest 10 events dominate the information in constraining the
Love-Q relation. Similar phenomena were also found in studies of constraining
the EOSs with GW observations~\cite{Lackey:2014fwa, Landry:2020vaw,
Pang:2020ilf, Finstad:2022oni, Bandopadhyay:2024zrr, Wang:2024xon}.

To our knowledge, we are the first to conduct a systematic study on different
parameterization models when inferring the Love-Q relation with GWs. Considering
the quartic polynomial model proposed by \citet{Yagi:2013awa} and the linear
model adopted by \citet{Samajdar:2020xrd} as two ends, we conduct the inference
with four polynomial models from linear to quartic terms. As shown in 
section~\ref{sec:results}, we quantitatively demonstrate that the linear model
is accurate enough to describe the Love-Q relation in the inference in the CE
and ET era. As the number of hyperparameters that parametrize the Love-Q
relation increases, more significant degeneracy or poorly constrained posteriors
appear, while the recovered Love-Q relations only keep similar in the region
where most data points gather.

We also test the potential of using the inferred Love-Q relation to constrain
modified gravity theories in section~\ref{sec:dCS}.  Taking the dCS gravity as
an example, we find that the inferred Love-Q relation can place a constraint on
the dCS characteristic lengthscale to about $\xi_{\mathrm{CS}}^{1/4} \lesssim
10$\,km, which is seven orders of magnitude tighter than that from the current
Solar System observations~\cite{Ali-Haimoud:2011zme, Yagi:2012ya}, and is
consistent with previous predictions with GWs~\cite{Yagi:2013bca, Yagi:2013awa}.
This highlights the power of inferring the Love-Q relation from GWs in testing
gravity theories in the strong-field regime. 

There can be several extensions to this study. Firstly, we assume  BNS systems
with aligned spins, while realistic BNS systems may have tilted spins and spin
precession, which leads to additional contribution in the GW
waveform~\cite{Williamson:2017evr, Purrer:2019jcp, Gamba:2020wgg}. The inaccuracy of 
waveform itself can also introduce a theoretical error to the posteriors of single event 
parameters, especially for events with high SNRs~\cite{Cutler:2007mi}.
Additionally, in the XG era, the GW signals are possibly overlapping with each
other, complicating the data analysis~\cite{Pizzati:2021apa, Samajdar:2021egv,
Wang:2023ldq, Johnson:2024foj, Wang:2025ckw}.  Moreover, the Yagi-Yunes
universal relations were originally found and discussed for a single NS under the slow-rotation assumption. 
For NSs in BNS systems or with high spins, these universal relations may break down and new relations could emerge. 
Our methodology in principle could still work when new universal relations do exist, but a careful inspection is required~\cite{Shao:2022koz,
Saffer:2021gak, Doneva:2013rha, Pappas:2013naa, Chakrabarti:2013tca, Yagi:2014bxa, Doneva:2014faa, Williams:2026jqv}. The octupole or higher-order multipole moments of NSs also
contribute to the GW waveform and can be universally related to the quadrupole
moment~\cite{Yagi_2017, Abac:2023ujg}. We leave the exploration of simultaneous
or independent inference of more universal relations to future studies.

%=============================
\acknowledgments

We thank the anonymous referee for helpful comments.
This work was supported by the Beijing Natural Science Foundation (QY25102,
1242018), the National Natural Science Foundation of China (123B2043, 12573042),
the National SKA Program of China (2020SKA0120300), the Max Planck Partner Group
Program funded by the Max Planck Society, and the High-Performance Computing
Platform of Peking University. 

%=============================
\appendix
\section{Dependence on the Number of Events}
\label{sec: appendix number of events}
%=============================

\begin{figure}[H]
    \centering
    \includegraphics[width=\textwidth]{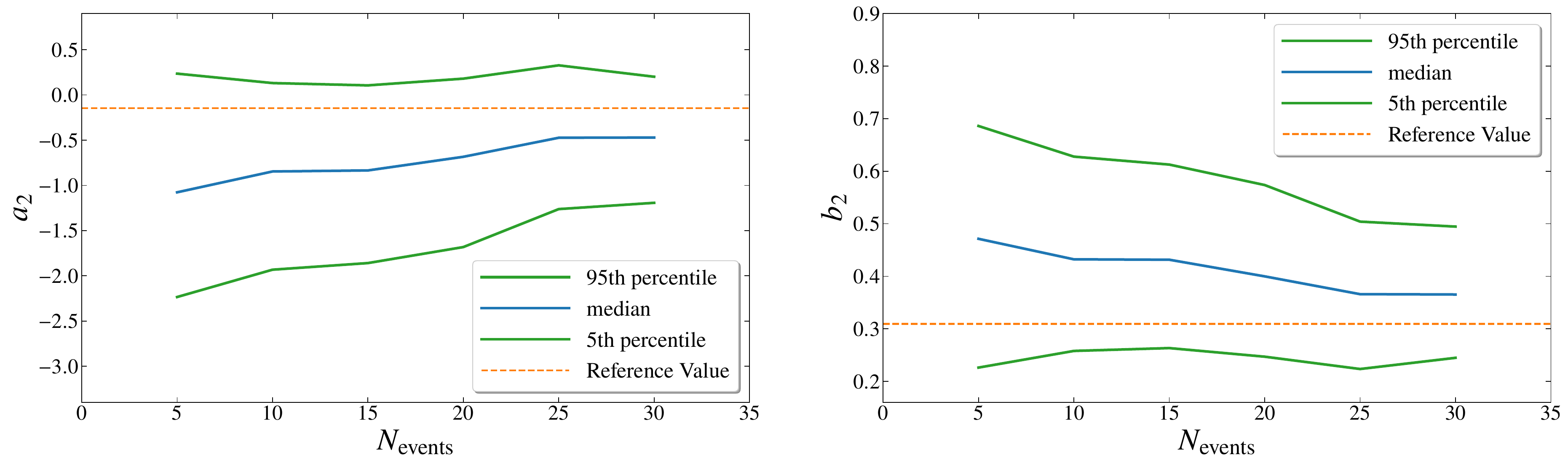}
    \caption{The dependency of the 5th, 50th and 95th percentiles of the hyperparameter posterior samples on the number of events. The orange dashed line represents the reference values, i.e. the linear fitting values obtained in section 2.1.}
    \label{fig:a_b_n}
\end{figure}

\begin{figure}[H]
    \centering
    \includegraphics[width=0.6\textwidth]{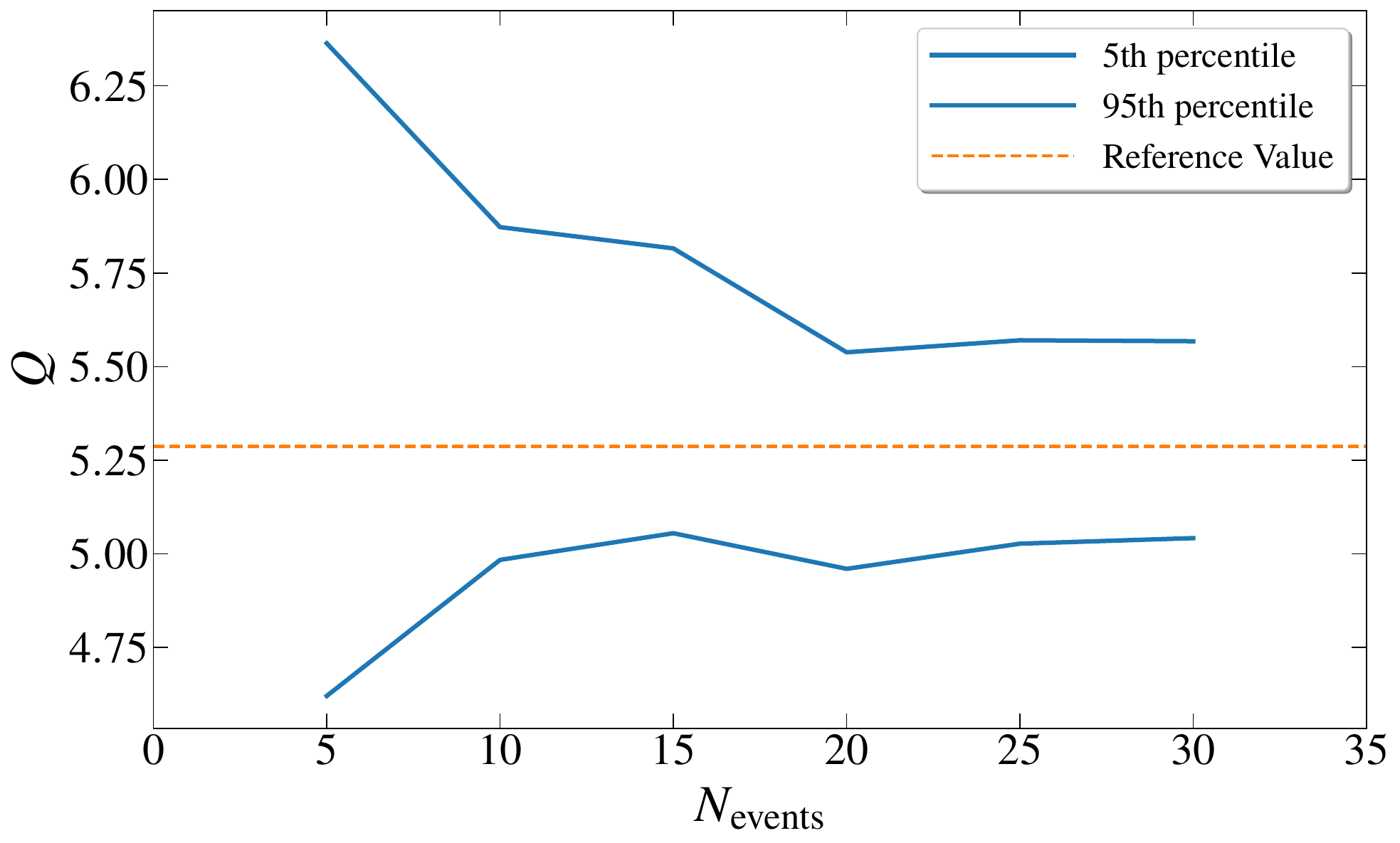}
    \caption{The dependency of 5th and 95th percentiles of $Q$ samples computed at $\Lambda=350$ (where most of the data points gather) on the number of events. The orange dashed line represents $Q$ calculated from the Yagi-Yunes relation.}
    \label{fig:width_n}
\end{figure}

In the main text, we have selected the loudest 20 events from the 1000 simulated
events in the inference.
Here we vary the number of events $N_{\text{events}}$ to see how the results
change. As shown in
figure~\ref{fig:a_b_n}, for a small $N_{\text{events}}$ (like $N=5$), there is a
noticeable offset between the median estimate and the reference value 
of the hyperparameters.
However, as $N_{\text{events}}$ increases, both the median and the credible
intervals shrink towards the reference values. 
to the reference values. Due to the decreasing SNRs of the
latter added samples, their contribution to the inference becomes less and less
significant, and the results keep almost unchanged.
The choice of 20 loudest
events can be justified with figure~\ref{fig:width_n}, where we observe that
the constraint to Love-Q relation at $\Lambda \sim 350$ (the ``waist'' part
in figure~3 of the manuscript) becomes stable when $N_{\text{events}}$ is larger
than 20. Similar results are also found in previous studies of constraining the
EOSs with GWs, where the several
 loudest events dominate the information~\cite{Lackey:2014fwa}.

%---------------------------------------------------------------------
%---------------------------------------------------------------------

%---------------------------------------------------------------------
%---------------------------------------------------------------------

%=============================
\section{Results for SLy}
\label{sec: appendix results for SLy}
%=============================

\begin{figure}[H]
    \centering
    \includegraphics[width=0.49\textwidth]{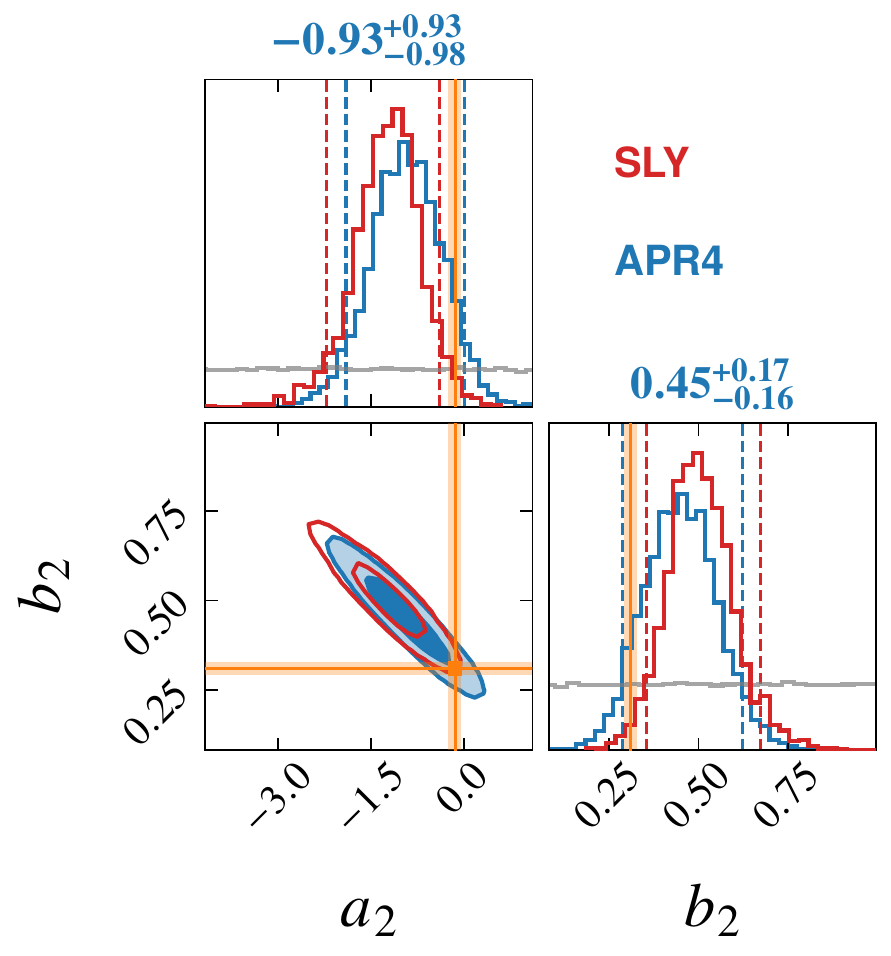}
    \caption{Corner plots for APR4 and SLy EOSs.}
    \label{fig:SLy_vs_APR4}
\end{figure}

\begin{figure}[H]
    \centering
    \includegraphics[width=\textwidth]{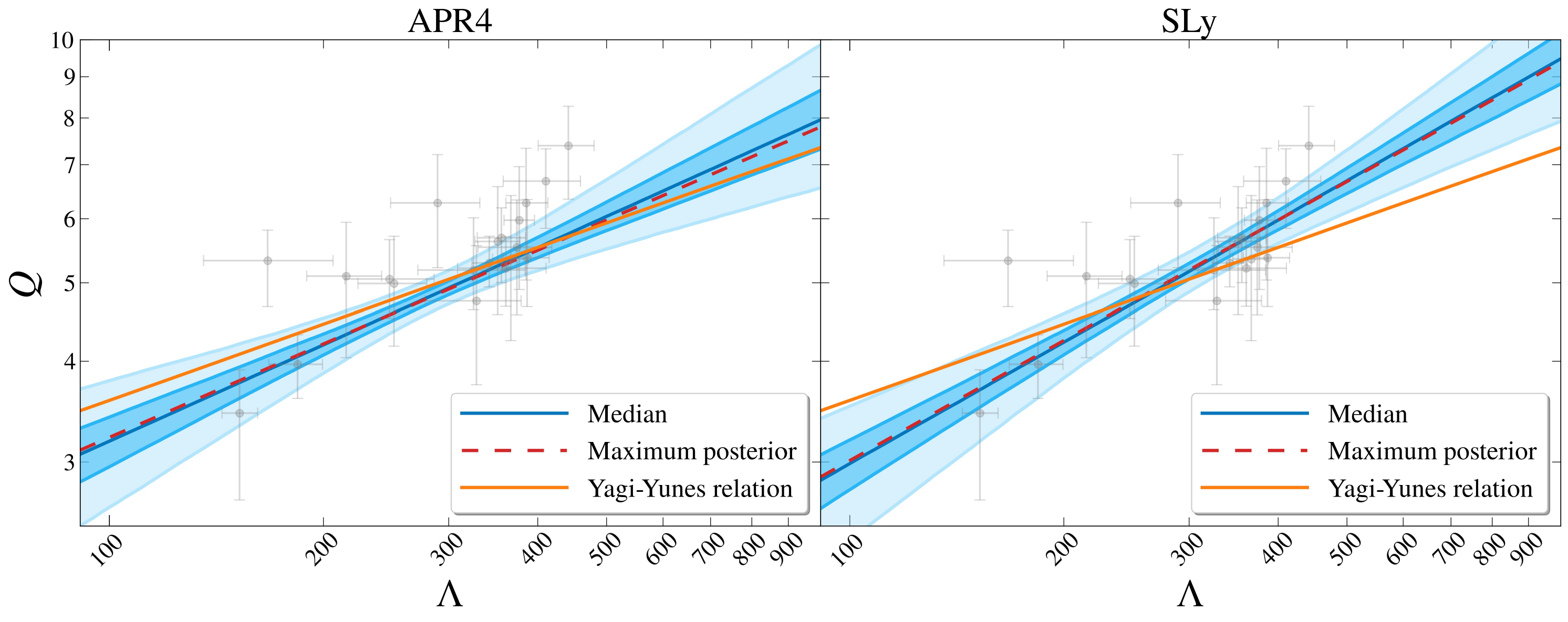}
    \caption{Constraints to Love-Q relation in a linear model for APR4 and SLy.}
    \label{fig:SLy}
\end{figure}

Different assumptions of EOS provide different combinations of $\Lambda$ and $Q$, 
which still follow the universal relation. As a supplement, we have repeated the inference of linear model for SLy, 
another EOS consistent with GW170817~\cite{LIGOScientific:2017vwq, LIGOScientific:2018cki, LIGOScientific:2018hze}.
The results are presented and compared with those of APR4 in figure~\ref{fig:SLy_vs_APR4} and figure~\ref{fig:SLy}.

As shown in figure~\ref{fig:SLy_vs_APR4}, 
the uncertainties of fitting coefficients brought by multiple EOSs are much smaller 
compared to those from the hierarchical inference. 
Thus the change of EOS would not introduce a significant change to the width of constraints, 
while the median may probably change with EOSs. And in figure~\ref{fig:SLy}. The median and maximum
posterior estimations change, but the width of 90\% credible region remains
insensitive to the choice of EOS.

%---------------------------------------------------------------------
%---------------------------------------------------------------------

%---------------------------------------------------------------------
%---------------------------------------------------------------------

\bibliographystyle{apsrev4-1}
\bibliography{HBAGW_jcap}
\end{document}